\documentclass[fleqn,usenatbib]{mnras}

\usepackage[T1]{fontenc}
\usepackage{ae,aecompl}
\usepackage[utf8]{inputenc}
\usepackage[english]{babel}
\usepackage{rotating}
\usepackage{tabularx,multirow}
\usepackage{url}
\usepackage{diagbox}
\usepackage{tikz,wrapfig}

\usepackage{latexsym}	
\usepackage{array}
\newcolumntype{P}[1]{>{\centering\arraybackslash}p{#1}}

\usepackage{mathtools}
\usepackage{xcolor}

\DeclareMathOperator*{\argmax}{arg\,max}

\usepackage{amsmath,bbm,bm,amssymb,mathtools,tikz}
\usepackage{soul}
\usepackage{lmodern}
\usepackage{graphicx,psfrag,epsf,natbib,subfig,wrapfig}
\usepackage{float}
\floatstyle{plaintop}
\restylefloat{table}
\usepackage{booktabs}
\usepackage{multirow}
\usepackage[export]{adjustbox}

\usepackage{verbatim}
\usepackage{booktabs}
\usepackage{xcolor}
\usepackage{tikz}
\usetikzlibrary{shapes.geometric, arrows}

\usepackage{algorithm}
\usepackage{algpseudocode}

\usepackage{hyperref}
\usepackage{url}

\newcommand*{\boldone}{\text{\usefont{U}{bbold}{m}{n}1}}

\newcommand{\bLambda}{\boldsymbol{\Lambda}}

\newcommand{\bSigma}{\boldsymbol{\Sigma}}
\newcommand{\bPsi}{\boldsymbol{\Psi}}

\newcommand{\bTheta}{\boldsymbol{\Theta}}
\newcommand{\bOmega}{\boldsymbol{\Omega}}

\newcommand{\bF}{\boldsymbol{F}}

\newcommand{\bI}{\boldsymbol{I}}

\newcommand{\bk}{\boldsymbol{k}}

\newcommand{\bq}{\boldsymbol{q}}

\newcommand{\bU}{\boldsymbol{U}}
\newcommand{\bx}{\boldsymbol{x}}
\newcommand{\bX}{\boldsymbol{X}}

\newcommand{\bZ}{\boldsymbol{Z}}

\renewcommand{\Pr}{{\mathbb P}\textnormal{r}}

\let\muold\mu
\renewcommand{\mu}{{\bm\muold}}
\let\LambdaOLD\Lambda
\renewcommand{\Lambda}{{\bm\LambdaOLD}}
\let\Psiold\Psi
\renewcommand{\Psi}{{\bm\Psiold}}
\let\Usiold\Upsilon
\renewcommand{\Upsilon}{{\bm\Usiold}}
\let\Sigmaold\Sigma
\renewcommand{\Sigma}{{\bm\Sigmaold}}
\let\Gammaold\Gamma
\renewcommand{\Gamma}{{\bm\Gammaold}}
\let\sigmaold\sigma
\renewcommand{\sigma}{{\bm\sigmaold}}

\newcommand{\normal}{{\mathcal N}}

\usepackage[many]{tcolorbox}
\newtcolorbox{mybox}[1][]{
    tikznode boxed title,
    enhanced,
    arc=0mm,
    interior style={white},
    attach boxed title to top center= {yshift=-\tcboxedtitleheight/2},
    fonttitle=\bfseries,
    colbacktitle=white,coltitle=black,
    boxed title style={size=normal,colframe=white,boxrule=0pt},
    #1}

\definecolor{gamaopt4}{RGB}
{103,     0,   31}
\definecolor{gamaopt3}{RGB}
{178,    24,   43}
\definecolor{gamaopt6}{RGB}
{214,    96,   77}
\definecolor{gamaopt7}{RGB}
{244,   165,  130}
\definecolor{gamaopt1}{RGB}
{146,   197,  222}
\definecolor{gamaopt8}{RGB}
{67,   147,  195}
\definecolor{gamaopt2}{RGB}
{33,   102,  172}
\definecolor{gamaopt5}{RGB}
{5,    48,   97}

\definecolor{gamaopt.no34}{RGB}{95,   92,  132}

\definecolor{gamaopt.34}{RGB}{138,    11,   36}
\tikzstyle{gopt1} = [ellipse, minimum width=1cm, minimum height=0.5cm,text centered, 
draw=black,
line width=0.5pt,
fill=gamaopt1]
\tikzstyle{gopt2} = [ellipse, minimum width=1cm, minimum height=0.5cm,text centered, draw=black,
line width=0.5pt, fill=gamaopt2]
\tikzstyle{gopt3} = [ellipse, minimum width=1cm, minimum height=0.5cm,text centered, draw=black,
line width=0.5pt, fill=gamaopt3]
\tikzstyle{gopt4} = [ellipse, minimum width=1cm, minimum height=0.5cm,text centered, draw=black,
line width=0.5pt, fill=gamaopt4]
\tikzstyle{gopt5} = [ellipse, minimum width=1cm, minimum height=0.5cm,text centered, draw=black,
line width=0.5pt, fill=gamaopt5]
\tikzstyle{gopt6} = [ellipse, minimum width=1cm, minimum height=0.5cm,text centered, draw=black,
line width=0.5pt, fill=gamaopt6]
\tikzstyle{gopt7} = [ellipse, minimum width=1cm, minimum height=0.5cm,text centered, draw=black,
line width=0.5pt, fill=gamaopt7]
\tikzstyle{gopt8} = [ellipse, minimum width=1cm, minimum height=0.5cm,text centered, draw=black,
line width=0.5pt, fill=gamaopt8]

\tikzstyle{startstop} = [rectangle, rounded corners, minimum width=3cm, minimum height=1cm,text centered, draw=black, fill=red!30]

\tikzstyle{gopt34m1} = [rectangle, rounded corners, minimum width=1.25cm, minimum height=1cm,text centered, draw=black,
line width=0.5pt, fill=gamaopt.34]

\tikzstyle{goptno34} = [rectangle, rounded corners, minimum width=2.5cm, minimum height=1.75cm, text centered, draw=black,
line width=0.5pt, fill=gamaopt.no34]

\tikzstyle{arrow} = [thick,->,>=stealth]

\title[Clustering and characterisation of GAMA data]{Multi-layered model-based characterisation of the local-Universe galaxy data from the GAMA survey}

\author[Dai, Maitra and Baldry]{
  Fan Dai,$^{1}$\thanks{E-mail: fan.dai@ndsu.edu (FD)}
  Ranjan Maitra,$^{2}$\thanks{E-mail: maitra@iastate.edu (RM)}
  and Ivan K. Baldry$^{3}$\thanks{E-mail: i.baldry@ljmu.ac.uk (IB)}\\
$^{1}$ Department of Statistics, North Dakota State University, 1340 Administration Ave, Fargo, ND 58105, USA\\
$^{2}$ Department of Statistics, Iowa State University, 2438, Osborn Drive, Ames, Iowa 50011-1090, USA\\
$^{3}$ Astrophysics Research Institute, Liverpool John Moores University, IC2, Liverpool Science Park, 146 Brownlow Hill, Liverpool L3 5RF, UK
}
\date{Accepted 2026 May 21. Received 2026 May 20; in original form 2026 April 21}

\pubyear{2026}

\begin{document}
\label{firstpage}
\pagerange{\pageref{firstpage}--\pageref{lastpage}}
\maketitle

\begin{abstract}
Understanding the formation and evolution of galaxy populations requires robust classification and characterisation techniques that jointly account for internal galaxy properties and environment. We analyse $5,306$ galaxies from the Galaxy And Mass Assembly (GAMA) survey, described by stellar mass, specific star formation rate, $u-r$ colour, half-light radius, S\'ersic index, and a combined environmental measure given by the optimal density. Unlike distance-based unsupervised clustering methods, our framework provides a probabilistic characterisation of galaxy populations, accommodates heavy-tailed feature distributions, and captures dependence among observables through latent factors. We model the sample using a $t$-mixture of factor analysers with group-specific latent structures (M$t$FAD), and then apply model-estimated overlap-based syncytial clustering (MOBSynC) to merge weakly separated groups and recover higher-level population structure. The first stage identifies eight simple clusters. The third and the fourth groups lie on the red, low-star-forming sequence and correspond to environmentally quenched and mass-quenched systems, respectively, while the sixth group traces the massive end of the star-forming sequence, and  the seventh group appears to represent a more heterogeneous population that may include transition objects. The remaining groups populate the low- to intermediate-mass blue sequence, including both compact and more extended star-forming galaxies. The second MOBSynC stage  merges the simple clusters into two compound groups: a red sequence formed by the third and the fourth groups, and the rest merging to form a broad blue sequence. Our results show that the familiar red-blue bimodality of local galaxies contains additional physically meaningful substructure linked to quenching pathway, morphology, and environment.

\end{abstract}




\section{Introduction}
\label{sec:intro}
Identifying and distinguishing diverse galaxies in the local Universe has long been of  major interest in astrophysics, providing key insight into the formation and evolution of galaxy populations under the influence of their inhabited environments \citep{Postmanetal1984,Mooreetal1995,Naabetal2001,parketal2007,kelvinetal2014,turneretal19}. Traditional approaches for classifying the local-Universe galaxies rely on predefined morphological schemes, such as the Hubble sequence \citep{hubble1926,sandage2005}, that broadly separate galaxies into disc-dominated and spheroid-dominated systems, followed by further characterization using individual astrophysical properties including star formation rates \citep{Smethurst2015}, galaxy colours \citep{kelvinetal2014}, stellar masses \citep{baldry2006,wel2014}, and S\'ersic indices over time \citep{lange2014}. While informative, these approaches often treat galaxy properties in isolation or impose rigid boundaries that may obscure more complex, multi-dimensional population structure; more importantly, with the rapidly increasing size and dimensionality of galaxy samples, such descriptive classification schemes become increasingly inadequate and impractical. In contrast, studies highlighting the joint influence of mass and environment—such as the separation of mass-driven and environment-driven quenching processes identified by \citet{peng2010}—motivate the use of clustering-based methods that can simultaneously integrate multiple correlated features and reveal latent groupings of galaxies shaped by both intrinsic properties and environmental effects.

Advanced statistical techniques have therefore been adopted to improve the classification of large and complex galaxy samples, moving beyond traditional, descriptive schemes toward data-driven approaches capable of handling multidimensional feature spaces. 
Existing literature on galaxy classification has focused on supervised learning approaches built on visually labelled training samples. For example, \cite{ball2004} utilizes the supervised artificial neural networks with Hubble-type labelled samples to classify $104619$ galaxies from the Sloan Digital Sky Survey (SDSS), while \cite{Aguerri2010} conducts the morphological classification of around $70,000$ galaxies from the SDSS DR7 spectroscopic sample using algorithms trained with visual classification results and \cite{Gravet2015} applies the convolutional neural networks to classify about $50,000$ galaxies and again, based on training samples that are visually classified.

Supervised learning methods are capable of distinguishing massive samples; however, by definition, they require labelled observations to train the classification algorithms and consequently, are inapplicable to cases with no existing labels, as often arises in many scientific studies. Identifying groups of galaxies, for instance, in the context of the studies in this paper, is done by cluster analysis, an unsupervised learning tool that has many different approaches and algorithms. One common technique is hierarchical clustering that builds a hierarchy of clusters based on dissimilarity measures between sets of observations, applied in \cite{ellisetal2005} that identifies two (early and late) types of galaxies from the Millennium Galaxy Catalogue. Another extremely common approach is $k$-means clustering, that iteratively assigns observations to $k$ clusters based on the nearest centroid (the mean of all the data points within the cluster). The $k$-means algorithm was used by \cite{Sanchezetal2010} to identify major and minor classes of all the galaxy spectra in the seventh and final SDSS data release, and \cite{turneretal19} to cluster around $7,000$ galaxies from the Galaxy And Mass Assembly (GAMA)
survey. Other unsupervised approaches have also been used to organise or represent complex galaxy populations, including self-organizing maps (SOM), which display similarities among galaxies in a multidimensional feature space using a two-dimensional representation~\citep{holwerda2022}; manifold-based representations for morphological classification~\citep{cooray2023}; and the Fisher expectation-maximization algorithm, which has been used to distinguish galaxies based on magnitudes and spectroscopic redshifts~\citep{Siudeketal2018}. These clustering methods, relying on measures of similarities or distance between sample points, while easy to implement, are unable to fully describe the underlying distributions of the grouped data. 

In contrast, model-based clustering (MBC) \citep[see, for instance][]{anderson03,mclachlanandpeel00,mardiaetal06,melnykovandmaitra10,chattopadhyayandmaitra17,chattopadhyayandmaitra18} is an attractive approach to clustering because it provides a principled probabilistic-based characterisation of groups in a dataset. Typically, the probabilistic model is a mixture of component distributions with parameters to capture the central tendency and variation within the group, each of which characterises different desired aspects of each group. MBC approaches have been applied to cluster galaxies in \cite{kelly2003} by means of a Gaussian mixture model (GMM) to separate around $3000$ galaxies in SDSS data, or in \cite{blacketal2024} where the GMM is used to characterise the red and blue sequences of DES galaxies in the COSMOS field. Separately, \cite{blacketal2022} proposed an error-corrected GMM developed in the space of broad-band optical colours across redshift for galaxy population characterization while \cite{zhangetal2023} modelled the conditional galaxy property distribution via the GMM. In each case, the GMM provides estimated ellipsoidal clusters of galaxies, which extends the isomorphic-cluster idea underlying $k$-means by providing a likelihood-based framework where each galaxy population is represented by a Gaussian distribution with its own mean and dispersion matrix. This yields ellipsoidal clusters, posterior membership probabilities, component-wise uncertainty estimates, and likelihood-based model comparison.

\begin{figure*}
    \centering
\includegraphics[width=\linewidth]{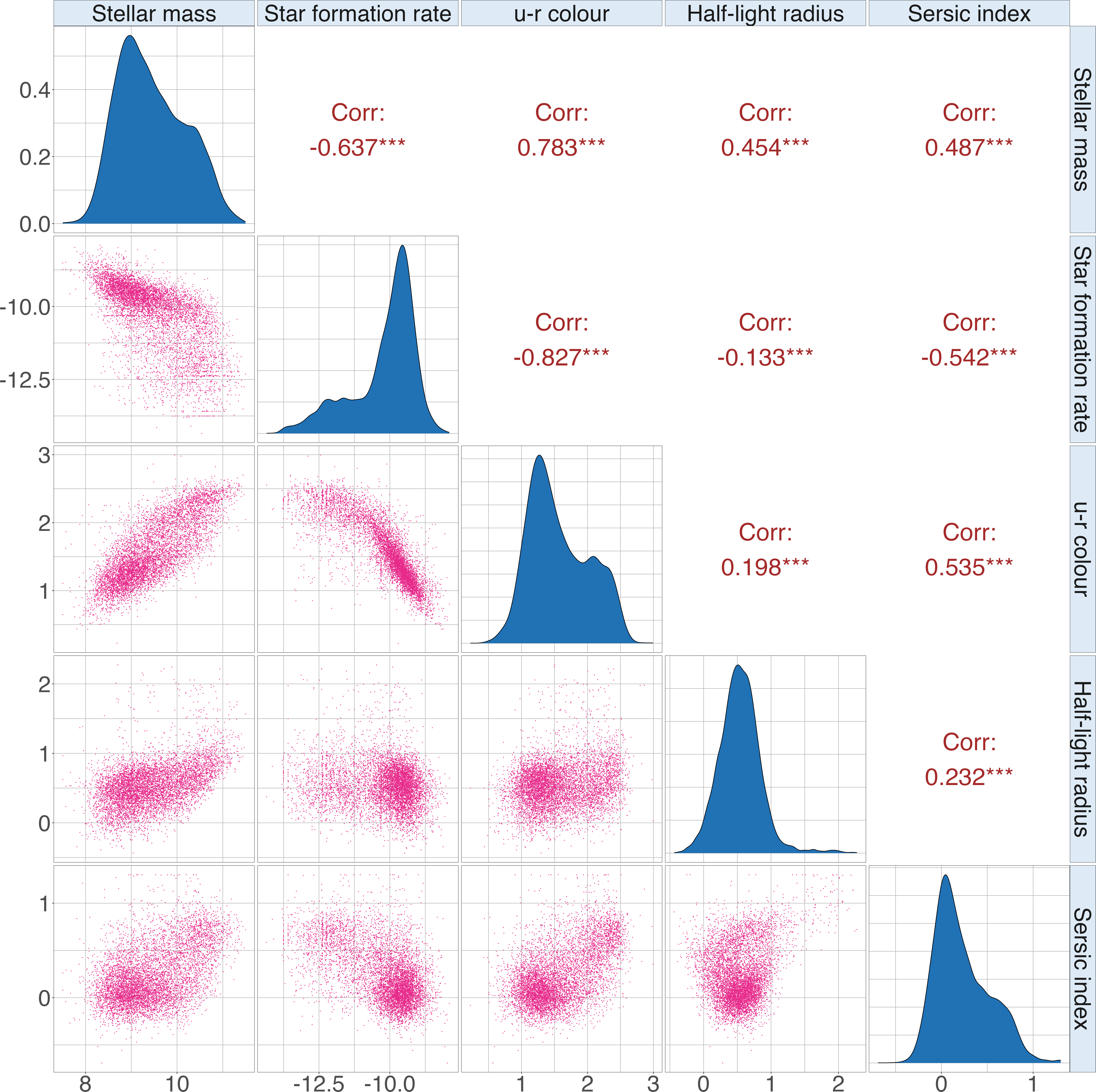}
    \caption{Densities and scatter plots of the five features (after $\log_{10}$ transformation except the $u-r$ colour) for the $7,187$ local-Universe galaxies from the GAMA survey. 
    Correlations between features are shown in the upper panel.}
    \label{fig:gama-feature}
\end{figure*}

A GMM is however less effective in describing samples with longer tails, as exhibited by most of the features (after $\log_{10}$ transformation except the $u-r$ colour) of the $7,187$ galaxies (see Fig~\ref{fig:gama-feature}
of our dataset that is described in greater detail in Section~\ref{sec:data}). Moreover, further investigation is required into the interdependencies among galaxy features and their dependencies on the local environment. For instance, local environmental density has been found to correlate with galaxy colour~\citep{baldry2006,vanderBurg18,Reeves21,Bhambhani23} and star formation rate~\citep{Schaefer18,Barsanti2018,trussler19,vandesande21,Sotillo-Ramos21}. Further, the existence of larger clusters that could arise from poorly separated groups remains unclear. Addressing these open questions demands probabilistic, flexible clustering frameworks that can capture intrinsic variability among galaxies while accounting for environmental effects and potential hierarchical organisation within galaxy populations.

In this paper,  we cluster and characterise the galaxy data using a generalised $t$-mixture of factor analysers with variable numbers of factors (M$t$FAD) developed by \cite{kareemanddai25}, plus a model-estimated overlap-based syncytial clustering (MOBSynC) adapted from \cite{almodovarandmaitra20,chattopadhyayetal22,daiandmaitra24}. The use of multivariate $t$-distributions offers a more robust modelling of the mixture components, and the factor analysers embedded in each $t$-distributed component can summarise all the data parameters using a few latent variables called factors, providing a better characterisation for the dispersion of the observations within the cluster. MOBSynC, on the other hand, uses measures of pairwise and generalised overlaps \citep{maitraandmelnykov10,melnykovandmaitra11,melnykovetal12} to create compound or composite groups by merging the simple clusters obtained from M$t$FAD. We apply the methods to the galaxy samples described in Section~\ref{sec:data}, and identify eight simple clusters characterised by distinct sets of factors, and two major clusters of the red and blue sequences, where the merged blue sequence is further described via its underlying factors.

The remainder of this article is organised as follows. In Section~\ref{sec:meth}, we introduce the M$t$FAD algorithm and the MOBSynC procedure. Section~\ref{sec:data}  describes the galaxy dataset that is analysed in Section \ref{sec:app}. Finally, Section~\ref{sec:con}
summarises the paper and discusses possible avenues for further work. 

\renewcommand{\hat}{\widehat}
\section{A multi-layered characterisation framework}
\label{sec:meth}
In this section, we present the statistical methodology underlying the clustering analysis used in this paper. Our approach begins with model-based clustering of the data using a $t$-mixture of group-specific factor analysers, together with efficient computational procedures for model parameter estimation. Building on the resulting initial partition, we then adapt the overlap-based merging framework of \citet{chattopadhyayetal22,daiandmaitra24} to combine clusters according to pairwise and generalized overlap measures. The proposed methods are implemented in our {\tt M$t$FAD} ($t$-Mixture of Factor Analysers in Data) and {\tt MOBSynC} programs, both written in the open-source statistical software {\tt R} \citep{R} and available at \url{https://github.com/fanstats/MBC-GAMA}.

\subsection{Clustering with \textit{t}-mixtures}
\label{sec:gmm}
A $t$-mixture model ($t$MM; \citealp{chattopadhyayandmaitra18}) has the same general mixture-model structure as a GMM, but replaces the Gaussian components with $p$-variate $t$ distributions, denoted by $t_p(\bmu,\bSigma;\nu)$, with mean vector $\bmu$, scale matrix $\bSigma$, and degrees of freedom $\nu$ that allows each component to accommodate heavier tails and potential outliers more robustly than a Gaussian distribution. Specifically, let $\bx_1,\bx_2,\dots,\bx_n$ be $p$-dimensional observations arising from a $t$MM with $K$ components, where $\bx_i$ belongs to the $k$th component with probability $\eta_k$, for $k=1,2,\dots,K$. Then, the observed data loglikelihood is
\begin{equation}\label{eqn:tmm-loglik}
\ell(\bTheta;\bX) = \sum_{i=1}^{n}\log\Big\{\sum_{k=1}^{K}\eta_{k}f_{t}(\bx_i;\bmu_k,\bSigma_k,\nu_k)\Big\},
\end{equation}
where $\bX {=} [\bx_1\vdots\bx_2\vdots\dots\vdots\bx_n]^\top$ is the data matrix, $\bTheta {=} \{(\eta_k,\bmu_k,\bSigma_k), k {=} 1,2,\ldots, K\}$ is the set of parameters characterizing the mixture model, and $f_t(\cdot)$ denotes the multivariate $t$ probability density function (PDF) for the $k$th mixture component.

For model parameter estimation, direct maximisation of Eq.~\eqref{eqn:tmm-loglik} is generally intractable, but can be carried out using the expectation-maximisation (EM) algorithm \citep{dempsteretal77,rubinandthayer82,mclachlanandkrishnan08}. To do so, an unobserved component indicator $z_i$ and a latent component-specific Gamma random variable $u_i$ are introduced for each observed $\bx_i$, with $\Pr(z_i=k)=\eta_k$ and the conditional distribution of $u_i$ given $z_i=k$ specified to be the Gamma $\mathcal{G}\left(\displaystyle\frac{\nu_k}{2},\frac{\nu_k}{2}\right)$ distribution. Then, conditional on $u_i$ and that $z_i=k$, $\bx_i$ is normally distributed as a $p$-variate  $\normal\left(\bmu_k,{\bSigma_k}/{u_i}\right)$ random vector. 
Then, the complete (or augmented) data loglikelihood for $t$MM is
\begin{equation}
\begin{split}
\label{eqn:tmm-emloglik}
\ell(\bTheta;\bX,\bZ,\bU)
&=\\
&\sum_{i=1}^{n}\sum_{k=1}^{K}\boldone(z_i=k)
\big\{\log \eta_k + 
\log\phi(\bx_i;\bmu_k,\frac{\bSigma_k}{u_i})\\
&+\log f_G(u_i;\frac{\nu_k}{2},\frac{\nu_k}{2})
\big\},
\end{split}
\end{equation}
where $\bZ=(z_1,z_2,\dots,z_n)$, $\bU=(u_1,u_2,\dots,u_n)$, $\boldone(\cdot)$ denotes the indicator function, $\phi(\cdot)$ and $f_G(\cdot)$ represent the Gaussian and Gamma density functions, respectively.

Starting from an initial value of $\bTheta$, and given $K$, the EM algorithm alternates between the E (or expectation)-step and the M (or maximisation)-step until convergence to a locally maximum likelihood (ML) solution, as outlined in Algorithm~\ref{alg1}. 

From the ML estimates obtained via EM, each observation is allocated to the $k$th component for which $\gamma_{ik}$ in Algorithm~\ref{alg1} is maximised. Having introduced the $t$-mixture model and its clustering framework, we now extend it to the $t$-mixture of factor analysers for richer characterisation of the component structure.

       \begin{algorithm}
 \caption{EM for $t$MM\label{alg1}}
 \begin{algorithmic}[1]
  \State Propose a set of initial model parameter values.
  \State Given current model parameters, compute the conditional expectations of $z_i$ and $u_i$ as follows.
  \[\gamma_{ik} = {\mathbb E}{[\boldone(z_i=k)|\mathbf{X}]} = \displaystyle{\frac{\eta_{k}f_{t}(\bx_i;\bmu_{k},\bSigma_{k},\nu_k)}{\sum_{k=1}^{K}\eta_{k}f_{t}(\bx_i;\bmu_{k},\bSigma_{k},\nu_k)}};\]
  \[\zeta_{ik} = {\mathbb E}{[u_i|\mathbf{X}]}=\displaystyle\frac{\nu_k  + p}{\nu_k + (\bx_i- \mu_k)^{\top} \Sigma_k^{-1}(\bx_i- \mu_k)}.\]

  \State Obtain model parameter estimates by maximising the expected complete data loglikelihood function, with updates computed as follows.
\begin{equation*}
\label{eq:est}
    \begin{split}
     &\eta_{k}^* =
        \displaystyle{\frac{\sum_{i=1}^{n}\gamma_{ik}}{n}},\\
     &\bmu_{k}^* =\frac{\sum_{i=1}^n \gamma_{ik} \zeta_{ik}  \bx_i}{\sum_{i=1}^n \gamma_{ik} \zeta_{ik} },\\
     &\bSigma_{k}^* = \displaystyle{\frac{\sum_{i=1}^n \gamma_{ik} \zeta_{ik} (\bx_i - \mu^*_k)(\bx_i - \mu^*_k)^{\top}}{\sum_{i=1}^n\gamma_{ik}\zeta_{ik}}},\\
     &\nu_k^* \text{: obtained numerically by solving the equation}\\
     &-\psi\left(\frac{\nu_k}{2}\right) + \log\left(\frac{\nu_k}{2} \right) + 1 + \frac{1}{n_k} 
\sum_{i=1}^n \gamma_{ik} \left(\log \zeta_{ik} - \eta_{ik}\right) \\
&+ \psi\left(\frac{\nu_k+p}{2}\right) - \log\left(\frac{\nu_k+p}{2}\right)  = 0.
    \end{split}
\end{equation*}

\State Iterate between Steps 2 and 3 until Eq.~\eqref{eqn:tmm-loglik} converges.
 \end{algorithmic} 
 \end{algorithm} 
The above formulation has provided the most general setup for the $t$MM. In many cases, the variability in each mixture component can be specified by a few unobservable factors that also serves to simplify the model by reducing the number of parameters. We introduce such a model next. 

\subsubsection{A \textit{t}MM with group-specific factor analysers}
\label{sec:fa}
We further characterise the $K$ components in the $t$MM by adopting a group-wise factor-analytic representation \citep{thurstone31,thurstone35,anderson03}, in which the $p$ observed variables in each ($k$th) group are explained by $q_k$ group-specific latent factors, with $q_k<\min(n,p)$ and $(p-q_k)^2>p+q_k$ for identifiability. Specifically, for the $k$th component,
\begin{equation}
\bSigma_k=\bLambda_k\bLambda_k^{\top}+\bPsi_k,    
\end{equation}
where $\bLambda_k$ is a $p\times q_k$ factor loading matrix whose \((j,l)\) entry represents the strength and direction of the relationship between the \(j\)th variable and the \(l\)th latent factor, where $j=1,2,\ldots, p$ and $l=1,2,\ldots,q$. $\bPsi_k$ is a diagonal matrix of feature-specific variances. Then, 
conditional on $u_i$ and $z_i=k$, 
$\bx_i$ can be modelled through a linear equation
\begin{equation}
\label{eq:fascores}
\bx_i =\bmu_k+\bLambda_k\bF_{i}+\bepsilon_{i},
\end{equation}
where given $u_i$ and that $z_i=k$, $\bF_i$ is $\normal\left(\boldsymbol{0},\bI_{q_k}/ u_i\right)$ distributed and  represents the $q_k$ latent factors, and conditionally independent of $\bepsilon_{i}$ that is $ \normal\left(\boldsymbol{0},{\bPsi_k}/{u_i}\right)$ distributed.

Similar to the setup in Section~\ref{sec:gmm},
 ML estimation of the model parameters may be carried out by combining the EM framework for $t$MM above with the classical EM approach for factor analysis \citep{dempsteretal77,Chen&Chen2008}. However, in practice, these iterative procedures may converge slowly, and are often sensitive to local maxima. In addition, standard formulations typically assume a common latent dimension (that is, $q_k{=}q$) across all $K$ components. To address these issues, we adopt the M$t$FAD algorithm of \citet{kareemanddai25} that extends the $t$-mixture of factor analysers to a more flexible version by allowing component-specific numbers of factors ($q_k$). The methodology incorporates (1) a profile likelihood strategy \citep{daietal20,daietal21} for efficient joint updating of $\bLambda_k$ and $\bPsi_k$ using matrix-free computations and (2) a stochastic initialization scheme~\citep{maitra13,gorenandmaitra22} to reduce sensitivity to local maxima. The main steps of M$t$FAD are summarised in Algorithm~\ref{alg2}.

       \begin{algorithm}
 \caption{M$t$FAD for $t$-mixture of group-specific factor analysers\label{alg2}}
 \begin{algorithmic}[1]
  \State Propose a set of initial model parameter values.
  \State Given current model parameters, compute the conditional expectations of $z_i$ and $u_i$ as follows.
  \[{\mathbb E}{[\boldone(z_i=k)|\mathbf{X}]} = \displaystyle{\frac{\eta_{k}f_{t}(\bx_i;\bmu_{k},\bLambda_k\bLambda_k^{\top} + \bPsi_k,\nu_k)}{\sum_{k=1}^{K}\eta_{k}f_t(\bx_i;\bmu_{k},\bLambda_k\bLambda_k^{\top} + \bPsi_k,\nu_k)}};\]
  \[{\mathbb E}{[u_i|\mathbf{X}]}=\displaystyle\frac{\nu_k  + p}{\nu_k + (\bx_i- \mu_k)^{\top} (\bLambda_k\bLambda_k^{\top} + \bPsi_k)^{-1}(\bx_i- \mu_k)},\]
  where $(\bLambda_k\bLambda_k^{\top} + \bPsi_k)^{-1}$ is evaluated using the Woodbury matrix identity~\citep{woodbury}.

  \State Update $\eta_{k},\bmu_{k}$, $\bSigma_{k}$, and $\nu_k$ as outlined in Algorithm \ref{alg1}.

\State Given $\bSigma_{k}$ and $q_k$, update $\bLambda_k,\bPsi_k$ as follows.
\begin{equation*}
    \begin{split}
      \bPsi_k^* = \argmax_{\bPsi_k} \mathrm{Q}_p(\bPsi_k), \quad \bLambda_k^* = h(\bPsi_k^*),
    \end{split}
\end{equation*}
where 
\begin{itemize}
    \item $\mathrm{Q}_p(\bPsi_k)$ is the expected complete data loglikelihood obtained after profiling out $\bLambda$ via $h(\bPsi_k)$.

\item $h(\bPsi_k)$ is derived from the score equations and determined by the $q$ largest eigenvalue–eigenvector pairs of $\bPsi_k^{-1/2}\bSigma_k^*\bPsi_k^{-1/2}$~\citep{kareemanddai25}.
\end{itemize}

\State Repeat Steps 2, 3 and 4 until Eq.~\eqref{eqn:tmm-loglik} converges.
 \end{algorithmic} 
 \end{algorithm} 
 \subsubsection{Number of clusters and factors}
Our framework so far has assumed that $K$ and all the $q_k$s are known, a largely unrealistic scenario in most practical settings. We therefore choose $K$ and $q_k$ by using 
the Bayesian information criterion (BIC; \citealp{Schwarz1978}) calculated by running 
Algorithm~\ref{alg2} for each of the candidate $K$ and $q_k$s, and then determine the optimal values to be the set that yields the smallest BIC.

\subsection{The MOBSynC algorithm for general-shaped groups}
\label{sec:merge}
The $t$-mixture of factor analysers models data as a collection of ellipsoidally shaped groups characterised by latent factors. However, as pointed out, for example, by \citet{almodovarandmaitra20}, by \citet{chattopadhyayetal22} or by \citet{daiandmaitra24},  some weakly separated groups may in fact represent subgroups within a larger compound cluster. Indeed, \citet{chattopadhyayetal22} provided a MOBSynC algorithm for $t$MM clusters, while \citet{daiandmaitra24} developed a similar algorithm for groups obtained using a Gaussian mixture of factor analysers model. Here, we adapt the same overlap-based principle to examine the propensity for merging among the clusters identified by M$t$FAD by computing pairwise and generalized overlaps using the fitted multivariate $t$ component densities. The procedure is described in Algorithm~\ref{alg3}.


\begin{algorithm}
    \caption{MOBSynC for clusters from M$t$FAD\label{alg3}}
    \begin{algorithmic}[1]
     \State Given the $K$ estimated clusters from M$t$FAD, compute the $K(K-1)/2$ pairwise overlaps $\omega_{k_1,k_2}$ and the generalised overlap $\ddot\omega$ as follows.
     \begin{equation*}
         \begin{split}
          \omega_{k_1,k_2} = & 
          \Pr\left(
       \frac{\eta_{k_1}f_{t}(\bx_i|z_i=k_2;\bmu_{k_1},\bSigma_{k_1},\nu_{k_1})}
       {\eta_{k_2}f_{t}(\bx_i|z_i=k_2;\bmu_{k_2},\bSigma_{k_2},\nu_{k_2})}
       >1\right) + \\
       &\Pr\left(
       \frac{\eta_{k_2}f_{t}(\bx_i|z_i=k_1;\bmu_{k_2},\bSigma_{k_2},\nu_{k_2})}
       {\eta_{k_1}f_{t}(\bx_i|z_i=k_1;\bmu_{k_1},\bSigma_{k_1},\nu_{k_1})}
       >1\right),\\
       \ddot\omega = &(\lambda^*_{\bOmega}-1)/(K-1),
         \end{split}
     \end{equation*}
     where $\lambda^*_{\bOmega}$ is the largest eigenvalue of the $K\times K$ matrix $\bOmega$ with pairwise overlaps $\omega_{k_1,k_2}$ as its entries.
       \State Merge the $k_1$th and $k_2$th clusters if $\omega_{k_1,k_2}>\kappa\ddot\omega$, where $\kappa$ is a positive integer indicating merging reluctance, with a larger value indicating that fewer pairs are merged in this step. Selection of the $\kappa$ value was discussed in \cite{daiandmaitra24}.
       \State Compute the probability that $\bx_i$ from the $m_1$th compound cluster $\mathcal{C}_{m_1}$ is misclassified to the $m_2$th compound cluster $\mathcal{C}_{m_2}$ as follows.
       \[\omega_{m_2|m_1} = \Pr\left(
       \frac{\sum_{j\in\mathcal{C}_{m_2}}\eta_{j}f_{t}(\bx_i|z_i={m_1};\bmu_{j},\bSigma_{j},\nu_{j})}
       {\sum_{l\in\mathcal{C}_{m_1}}\eta_{l}f_{t}(\bx_i|z_i={m_1};\bmu_{l},\bSigma_{l},\nu_{l})}
       >1\right),\] 
       which can be approximated via the following Monte Carlo methods~\citep{chattopadhyayetal22}:
       \begin{enumerate}
       \item Generate random samples $\bx_1, \bx_2,\cdots,\bx_M$ ($M$ is set to be $10^6$ in Section~\ref{sec:app}) from the mixture distribution of $\mathcal{C}_{m_1}$, which is defined as \[\sum_{l\in\mathcal{C}_{m_1}}\eta^*_{l}f_{t}(\cdot;\bmu_{l},\bSigma_{l},\nu_{l}),\quad \eta^*_{l} = \eta_{l}/\sum_{h\in\mathcal{C}_{m_1}}\eta_{h}.\] 
       
       \item Estimate $\omega_{m_2|m_1}$ as
       \[
          \hat\omega_{m_2|m_1} = \displaystyle\frac{1}{M}\sum_{i=1}^{M}\boldone\left\{\frac{\sum_{j\in\mathcal{C}_{m_2}}\eta_{j}f_t(\bx_i;\bmu_{j},\bSigma_{j},\nu_{j})}
       {\sum_{l\in\mathcal{C}_{m_1}}\eta_{l}f_t(\bx_i;\bmu_{l},\bSigma_{l},\nu_{l})}>1\right\}.
       \]  
   \end{enumerate}
   
       \State Compute the pairwise overlap between $\mathcal{C}_{m_1}$ and $\mathcal{C}_{m_2}$ as $\omega_{m_1,m_2} = \hat\omega_{m_1|m_2}+\hat\omega_{m_2|m_1}$, and update the $\ddot{\omega}$ following the approach described in Step 1.
   
       \State Repeat the merging steps 2, 3 and 4 until the current $\ddot\omega$, or its change compared to the previous stage, is below $10^{-3}$.
     \end{algorithmic}
     \end{algorithm}
  The result of applying Algorithm~\ref{alg3} is that we can get a detailed multi-layered characterisation of the galaxies in the GAMA database. 

\renewcommand{\hat}{\widehat}
\section{Galaxy Samples}
\label{sec:data}
We are interested in clustering and characterising the galaxy data 
using the methods outlined in this paper. For this, we aim to use a similar set of data as used by \citet{turneretal19}.
Those data were taken from the GAMA survey
\citep{baldry18} by selecting a galaxy sample at $z<0.06$. 
We use the same five features as \citet{turneretal19}
but add one additional environmental feature. 
To do this, we use a slightly higher redshift range ($0.05 < z < 0.08$) that is better suited for environmental measurements. 
The dataset therefore consists of local-Universe galaxies ($0.05<z<0.08$) from 
GAMA DR4 (\citealp{driver22}; \url{https://gama-survey.org/dr4/data/cat}). 
This corresponds to $7,187$ objects after removing observations with incomplete features and outliers. 
Each local object is described by five astrophysical features that capture the essential properties associated with the formation and evolution process of galaxies, including stellar mass (in $\log_{10}{\mathrm{M}_\odot}$), specific star formation rate (in $\log_{10}{\mathrm{yr}^{-1}}$), $u-r$ colour (in mags), half-light radius (in $\log_{10}{\mathrm{kpc}}$) and S\'ersic index (in $\log_{10}{\mathrm{n}}$).
All of which, except the $u-r$ colour
(a logarithmic flux ratio), are analysed after $\log_{10}$ transformation because of the high skewness in their measurements.

The features were obtained from the tables called 
MagPhysv06 \citep{driver18}, 
StellarMassesPanChromv24 \citep{taylor11} (for rest-frame $u-r$), 
and BDModelsv05 \citep{casura2022}. 
Notably we have updated the structural fitting of galaxy profiles
to those obtained by \citet{casura2022} except we only use the 
single S\'ersic profile fits. 

As the formation and evolution of galaxies are strongly influenced by their surrounding environment, three key measurements of the local galaxy environment are considered in our cluster analysis. These data, obtained from the GAMA DR4 file server (EnvironmentMeasuresv06), include the surface density, cylindrical count, and adaptive Gaussian environment parameter. The descriptions of these environmental features are as follows:

\begin{itemize}
  \item 
\textit{Surface density $\Sigmaold$} - based on the distance to the
5th nearest neighbour among the density defining population in a
velocity cylinder of $+/-1000$ km/s, i.e. $\displaystyle\frac{5}{\pi d_5^2}$ \citep{Brough20,Bhambhani23}.
\item
\textit{Cylindrical count $\mathrm{CC}$} - measured as the number of (other) galaxies from the density
defining population within a cylinder of co-moving radius $1$ Mpc and a
velocity range of $+/-1000$ km/s. The overdensity is given by $N_{\mathrm{cyl}} / (\bar{n}_\mathrm{ref}V_\mathrm{cyl})$, where $\bar{n}_\mathrm{ref} = 0.00734$ Mpc$^{-3}$ is the average number density of the density defining population \citep{Brough20,Bhambhani23}.
\item
  \textit{Adaptive Gaussian environment parameter {\rm AGE}} - computed as $$\mathrm{AGE} = \frac{1}{\sqrt{2\pi}\sigmaold}\sum_{i}{\exp{\left\{-\frac{1}{2}\left(\frac{r_{a,i}^2}{\sigmaold^2} + \frac{r_{z,i}^2}{(\mathrm{AGEScale}\cdot\sigmaold)^2}\right)\right\}}},$$where $r_a$ and $r_z$ are the distances from the centre of the adaptive Gaussian ellipsoid in the plane of sky and along the line-of-sight in co-moving Mpc, respectively, $\sigmaold = 2 $ Mpc, and $\mathrm{AGEScale}$ is the adaptive scaling factor used to scale the value of $\sigmaold$ along the redshift axis by up to a
factor of 3 for the highest density environments to compensate for the "finger-of-God" effect. This parameter is equivalent to a weighted local volume density of galaxies, where closer galaxies receive more weight than more distant ones \citep{Brough20,Bhambhani23}.
\end{itemize}

Within the GAMA sample, $5,306$ galaxies possess complete measurements for the three environmental parameters. To summarise these effects, we adopt a combined environmental measure known as optimal density, proposed by \citet{Bhambhani23} and detailed below, which effectively captures the variation in the red galaxy fraction and provides a more reliable measurement of local environmental influence. We therefore use

\textit{Optimal density $\LambdaOLD$} - computed as a linear combination of the surface density, cylindrical count, and adaptive Gaussian environment parameter: 
\begin{equation}\label{eqn:optimal-density}
\log \LambdaOLD = \log \Sigmaold + \alpha \log \mathrm{CC} + \beta \log \mathrm{AGE},
\end{equation}
where $\log \mathrm{CC}=-1$ at $\mathrm{CC}=0$. $\alpha$ and $\beta$ are numerically determined to maximize the red fraction range of $\LambdaOLD$.  For our fully observed sample of $5,306$ galaxies, the best results are $\alpha = 0.76$ and $\beta = 1.41$, with the local optimal density $\LambdaOLD$ achieving the highest red fraction range ($0.469$) compared to $\Sigmaold$ ($0.397$), $\mathrm{CC}$ ($0.425$), and $\mathrm{AGE}$ ($0.391$).

\begin{figure}
    \centering
\includegraphics[width=1\linewidth]{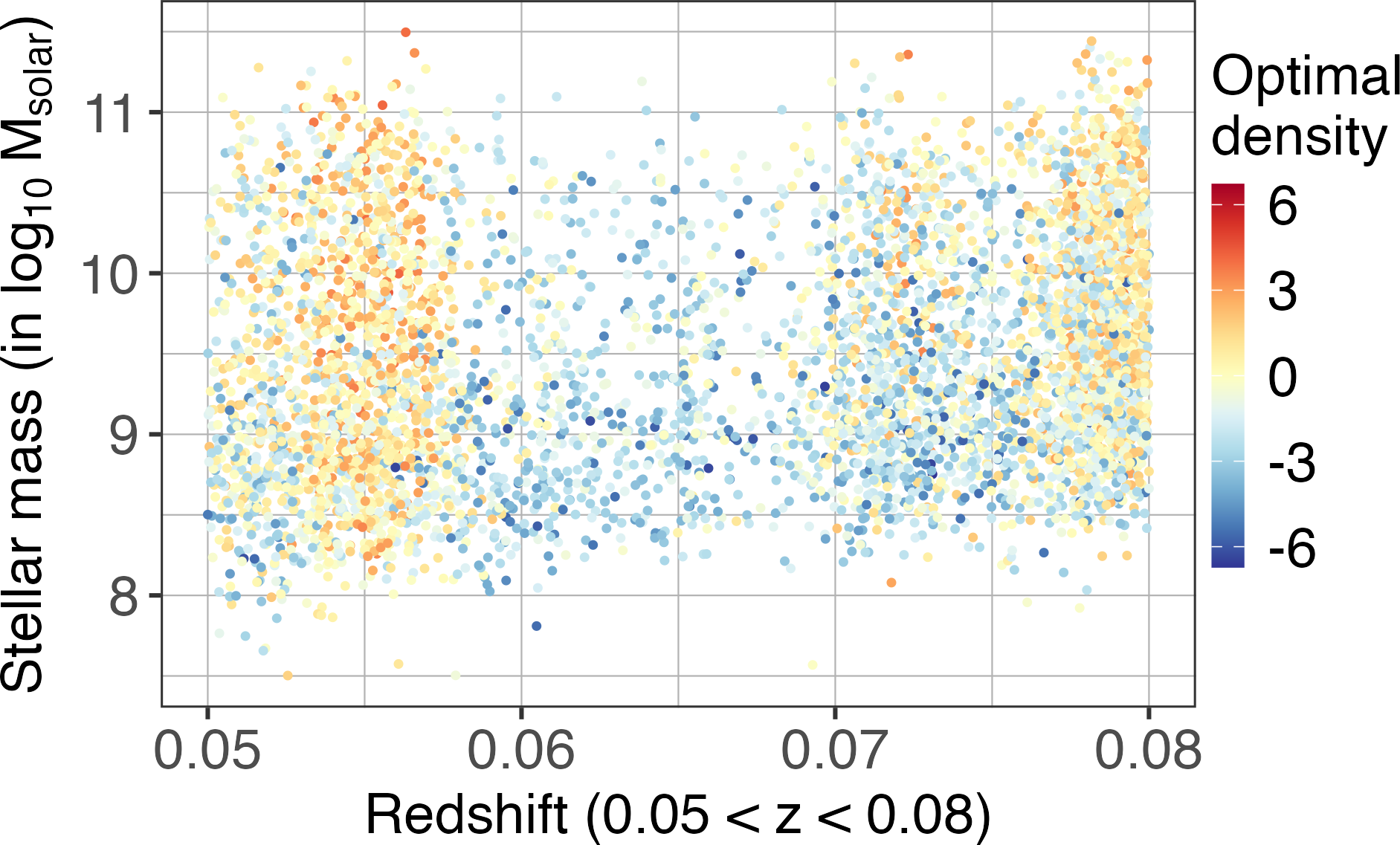}
    \caption{Redshift--stellar mass distribution of the $5,306$ GAMA galaxies with complete environmental measurements. Points are colour-coded by optimal density, the combined environmental measure. The sample is restricted to $0.05 < z < 0.08$.}
    \label{fig:mass-redshift}
\end{figure}

In sum, our final dataset consists of $5,306$ local-Universe galaxies with complete records on six parameters: Stellar mass, star formation rate, $u-r$ colour, half-light radius, S\'ersic index, and optimal density. Fig~\ref{fig:mass-redshift} shows that these galaxies lie within the redshift range $0.05<z<0.08$ and span approximately $7.5$--$11.5$ in $\log$ stellar masses. The colour gradient indicates that environmental density varies across the full sample, with higher optimal density values appearing more frequently among relatively massive galaxies.

\section{Results and Analysis}
\label{sec:app}
We applied the M$t$FAD algorithm to the $5,306$ fully observed GAMA galaxies, each described by five intrinsic astrophysical properties and one combined environmental variable—the optimal density. The modelling considered up to fifteen mixture components and up to two latent factors (consistent with the maximum permissible number of factors for six observed features). The resulting clustering structure and latent characterisations are summarised below.

\subsection{M\textit{t}FAD grouping}
\label{sec:simple-clusters}
\begin{figure}
    \centering
    \includegraphics[width=\linewidth]{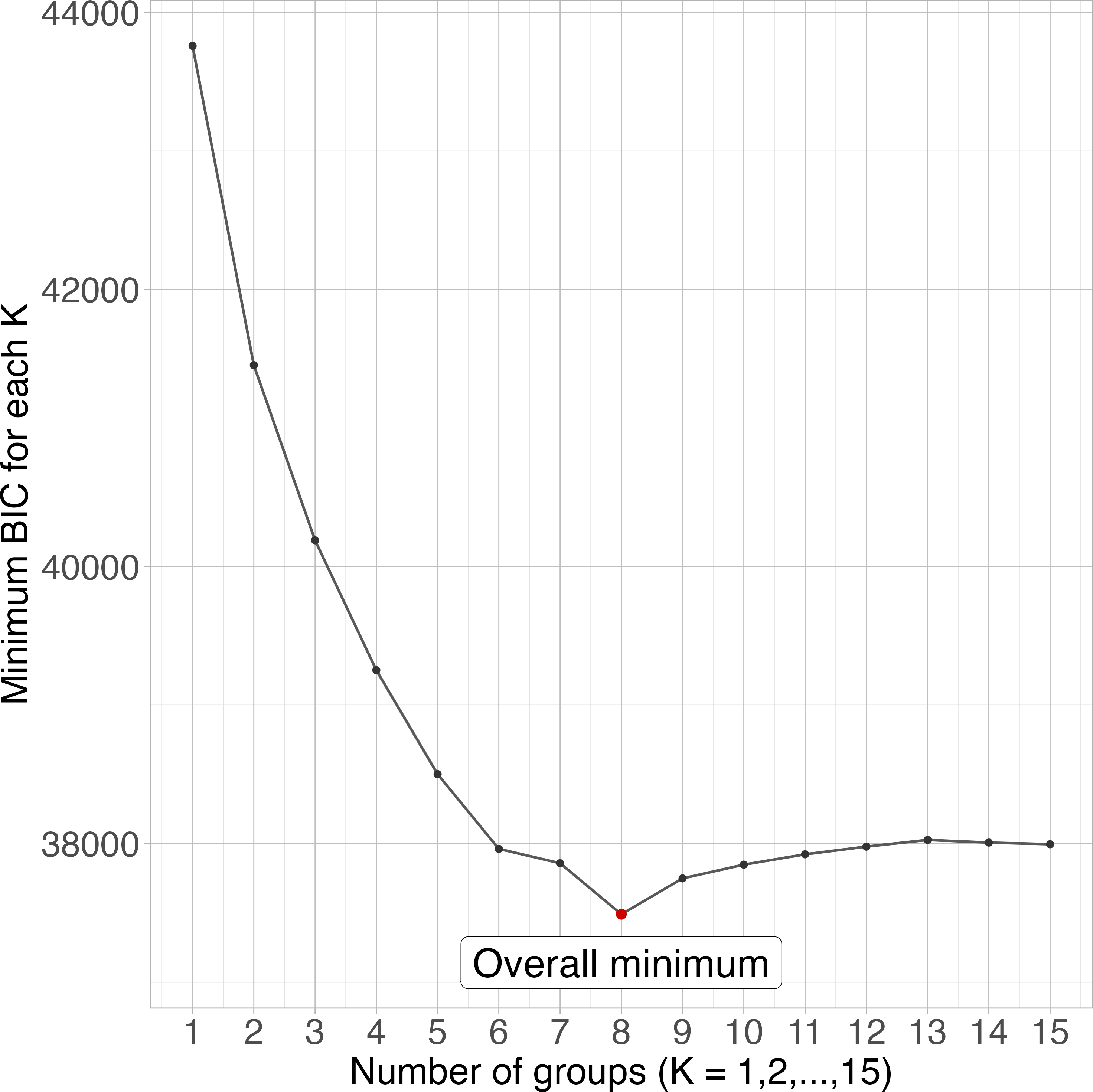}
    \caption{Minimum BIC values for the M$t$FAD model over candidate numbers of groups $K=1,2,\ldots,15$. For each fixed $K$, the plotted value is the smallest BIC obtained over all group-specific $q_k$ configurations, with $q_k\in\{1,2\}$ for $k=1,2,..,K$. The overall minimum is obtained at $K=8$.}
    \label{fig:gama-bic}
\end{figure}

\begin{figure*}
    \centering
   \includegraphics[width = \textwidth]{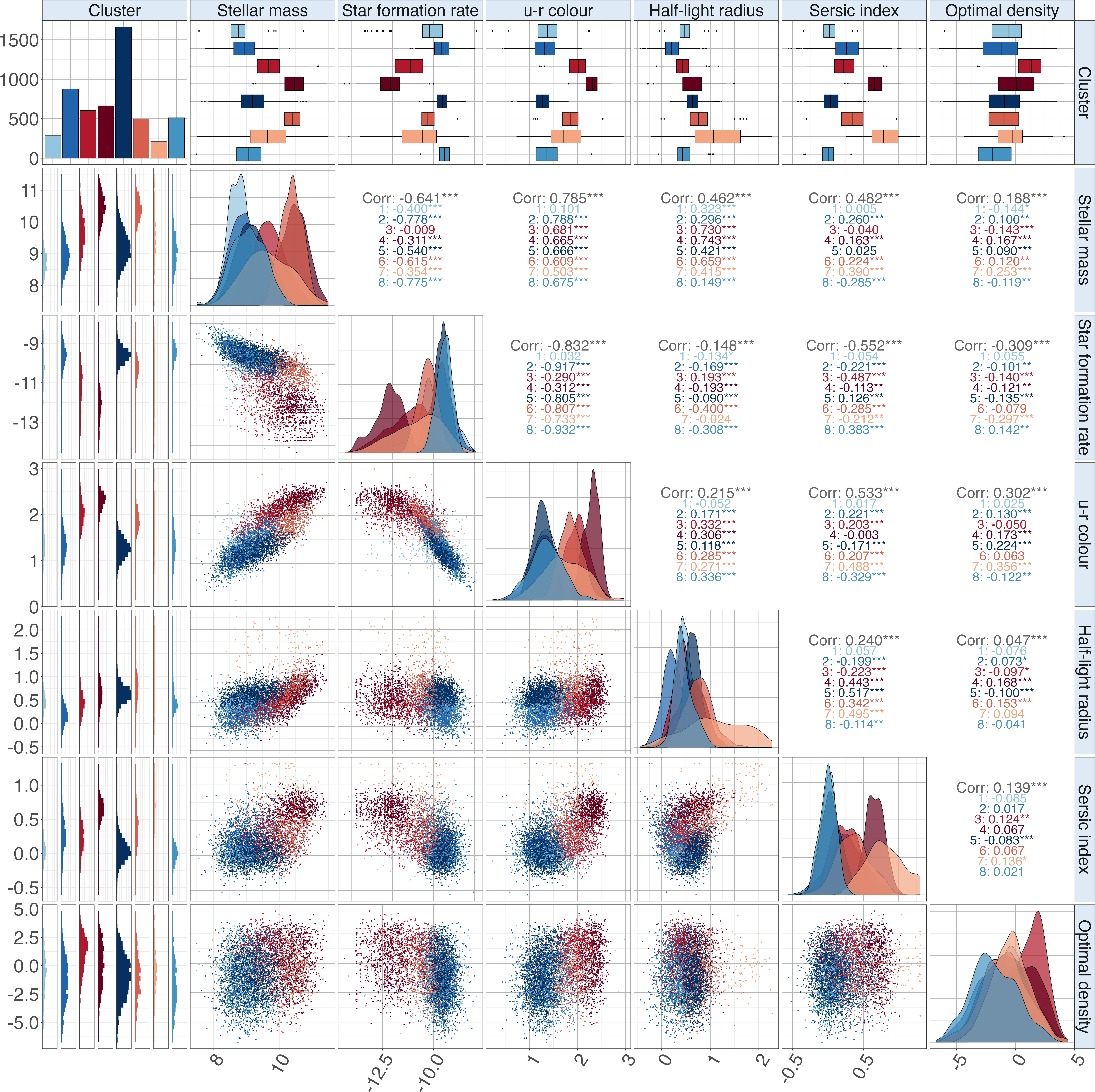}\\
    \centering
    \text{
     Simple cluster \tikz\draw[gamaopt1,fill=gamaopt1] (0,0) circle (.5ex); 1
     \tikz\draw[gamaopt2,fill=gamaopt2] (0,0) circle (.5ex); 2
     \tikz\draw[gamaopt3,fill=gamaopt3] (0,0) circle (.5ex); 3
     \tikz\draw[gamaopt4,fill=gamaopt4] (0,0) circle (.5ex); 4
     \tikz\draw[gamaopt5,fill=gamaopt5] (0,0) circle (.5ex); 5
     \tikz\draw[gamaopt6,fill=gamaopt6] (0,0) circle (.5ex); 6
     \tikz\draw[gamaopt7,fill=gamaopt7] (0,0) circle (.5ex); 7
     \tikz\draw[gamaopt8,fill=gamaopt8] (0,0) circle (.5ex); 8
    }
    \caption{Data with Optimal Density: Densities and scatter plots of the six features: Stellar mass, star formation rate, $u-r$ colour, half-light radius, S\'ersic index, and optimal density, for simple clusters (indicated by colours). Correlations between features are shown in the upper panel.}
    \label{fig:opt-initial-group-feature}
\end{figure*}

Our algorithm when applied with BIC selected $K = 8$ simple clusters (see Table~\ref{tab:gama-bic} and Fig~\ref{fig:gama-bic}).
\begin{table}
\centering
\caption{Model-selection results for M$t$FAD over candidate numbers of groups $K = 1, 2,\ldots, 15$. For each $K$, the minimum BIC is taken over all possible group-specific $q_k$ configurations $\mathbf q_K=(q_1,\ldots,q_K)$, with $q_k\in\{1,2\}$ for $k = 1, 2,\ldots, K$. Here $\Delta\mathrm{BIC}=\mathrm{BIC}_{K-1}-\mathrm{BIC}_{K}$ for $K\geq2$. The selected model is highlighted in bold.}
\label{tab:gama-bic}
\addtolength{\tabcolsep}{-0.35em}
\begin{tabular}{c|l|c|c}
\hline\hline
$\mathbf{K}$ & \textbf{Selected} $\mathbf q_K$ & \textbf{Min. BIC} & $\bm\Delta$\textbf{BIC} \\
\hline\hline
1  & $(2)$ & 43757.92 & -- \\
2  & $(2,2)$ & 41452.73 & 2305.19 \\
3  & $(2,2,2)$ & 40188.36 & 1264.37 \\
4  & $(2,2,2,2)$ & 39250.83 & 937.53 \\
5  & $(2,2,2,2,1)$ & 38500.60 & 750.23 \\
6  & $(2,2,2,2,1,2)$ & 37960.88 & 539.72 \\
7  & $(1,2,2,2,2,2,2)$ & 37857.39 & 103.49 \\
$\mathbf{8}$ & $\mathbf{(1,2,2,2,2,2,2,1)}$ & $\mathbf{37489.09}$ & $368.30$ \\
9  & $(2,1,2,2,2,2,2,2,1)$ & 37747.81 & -258.72 \\
10 & $(2,1,2,1,1,2,2,2,2,1)$ & 37847.07 & -99.26 \\
11 & $(2,1,1,2,2,2,2,2,1,2,1)$ & 37921.59 & -74.52 \\
12 & $(2,1,2,1,1,2,2,2,1,2,2,2)$ & 37977.30 & -55.71 \\
13 & $(1,2,2,2,1,2,2,2,2,2,2,2,1)$ & 38025.69 & -48.39 \\
14 & $(2,1,1,2,2,2,2,2,1,2,1,1,1,1)$ & 38006.58 & 19.11 \\
15 & $(2,1,2,2,2,1,2,2,2,2,2,2,1,2,1)$ & 37994.32 & 12.26 \\
\hline
\end{tabular}
\end{table}
The optimal number of factors for each group was $q_k = 1$ for Groups 1 and 8, and $q_k = 2$ for the remaining groups. We also see that Groups~2, 3, 5, 6 and 8 are essentially normally-distributed given that $\hat{\nu_k} > 100$ for these groups. Table~\ref{tab:simple-cluster-interpretation}
lists the galaxy counts of the eight estimated groups, where we see that Group~7 contains the fewest number of galaxies which is around $12\%$ of the largest Group~5 in terms of the sample size. Table~\ref{tab:gama-feature} presents the mean and standard deviation for each of the six parameters
within the identified groups, while Fig~\ref{fig:opt-initial-group-feature}
provides a visual summary of their distributional patterns. Groups~3 and 4 are characterised by comparatively higher stellar masses, redder $u-r$ colours, and lower average star formation rates, consistent with more evolved galaxy populations. Among them,  Group~4 is particularly distinguished by the reddest colours, the most suppressed star formation activity, and relatively large S\'ersic indices with less variations, while Group~3 stands out as the cluster with the highest optimal environmental density. In contrast, Groups~1, 2, 5, and 8 are generally associated with lower stellar masses, bluer colours, smaller S\'ersic indices, and higher average star formation rates. Among these, Groups~2 and~8 show the lowest optimal environmental densities. Groups~6 and 7 also exhibit relatively low optimal densities together with the largest half-light radii. In particular, Group~7 has the largest mean half-light radius and one of the highest S\'ersic indices, with the greatest variation in both quantities. 

\begin{figure*}
\centering
\mbox{
\subfloat{\includegraphics[width=0.5\textwidth]{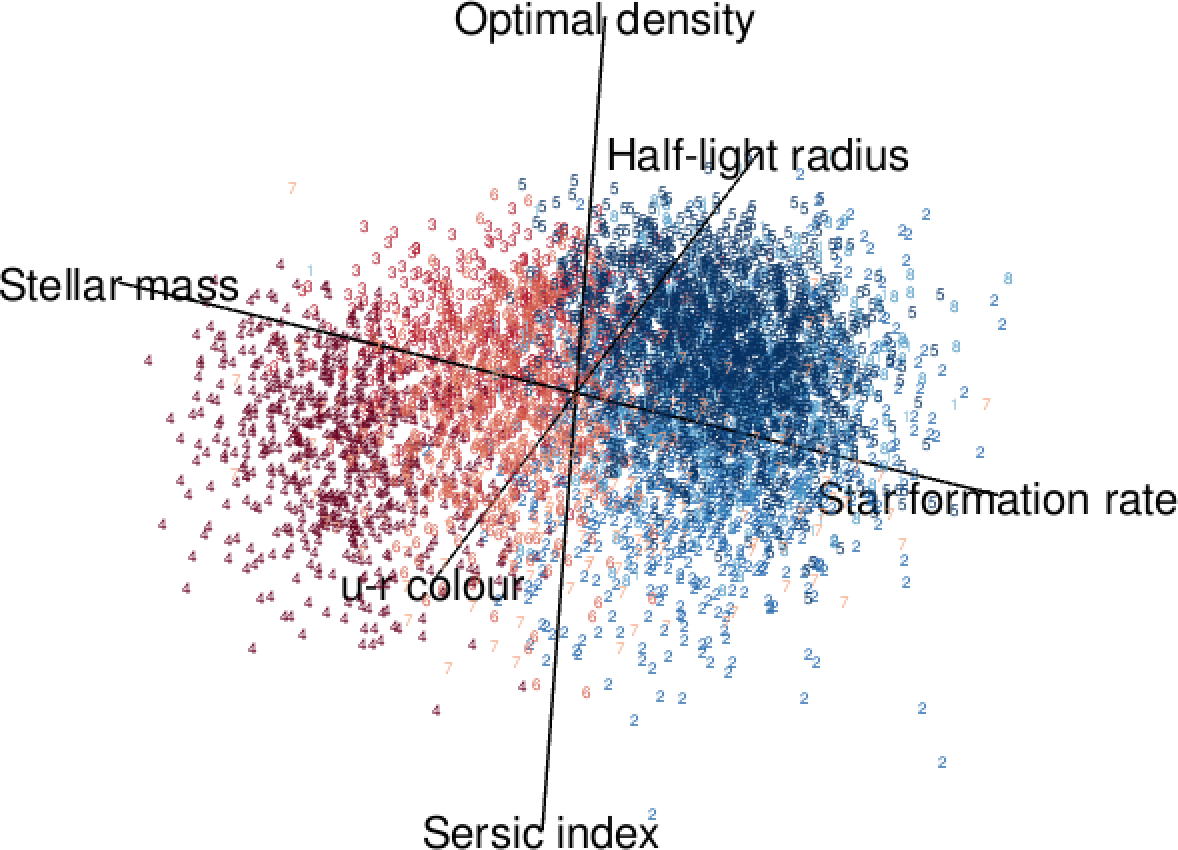}}
\subfloat{\includegraphics[width=0.5\textwidth]{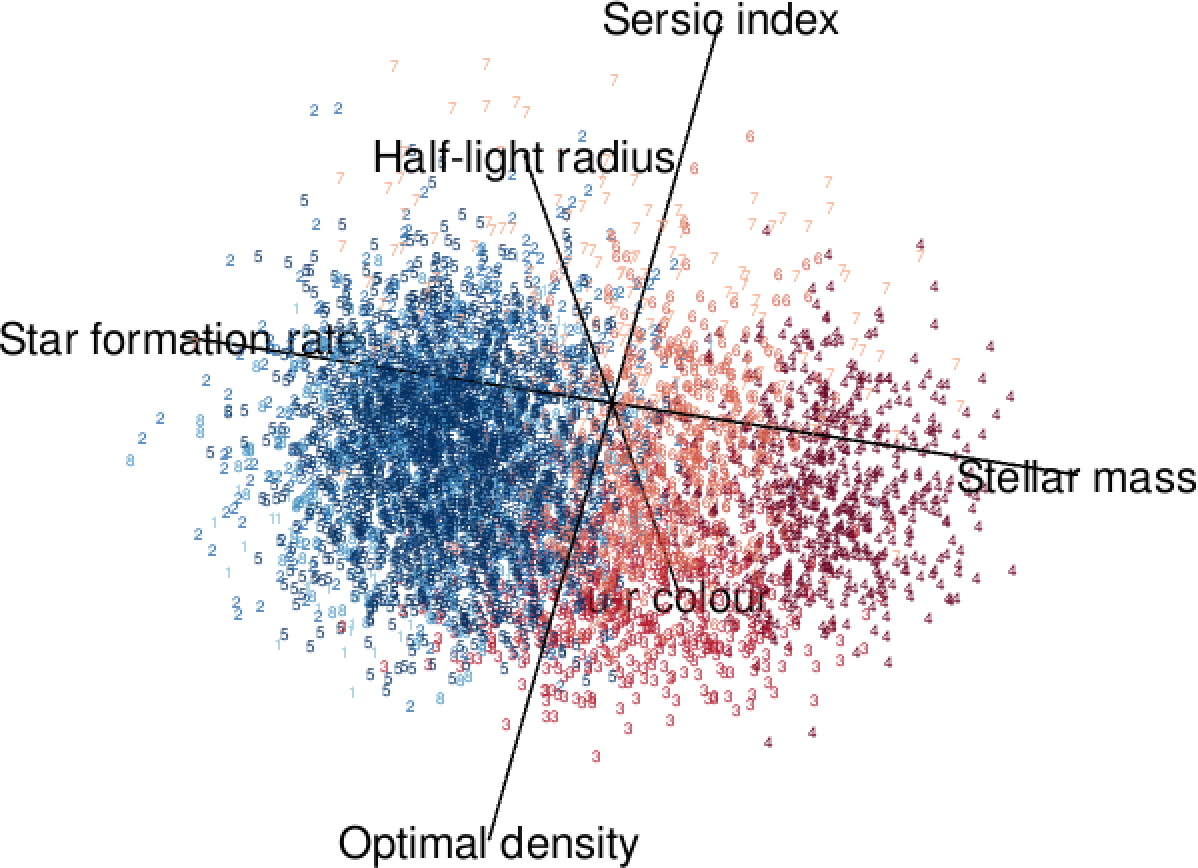}}
}
 \centering
    \text{
     Simple cluster \tikz\draw[gamaopt1,fill=gamaopt1] (0,0) circle (.5ex); 1
     \tikz\draw[gamaopt2,fill=gamaopt2] (0,0) circle (.5ex); 2
     \tikz\draw[gamaopt3,fill=gamaopt3] (0,0) circle (.5ex); 3
     \tikz\draw[gamaopt4,fill=gamaopt4] (0,0) circle (.5ex); 4
     \tikz\draw[gamaopt5,fill=gamaopt5] (0,0) circle (.5ex); 5
     \tikz\draw[gamaopt6,fill=gamaopt6] (0,0) circle (.5ex); 6
     \tikz\draw[gamaopt7,fill=gamaopt7] (0,0) circle (.5ex); 7
     \tikz\draw[gamaopt8,fill=gamaopt8] (0,0) circle (.5ex); 8
    }
 \caption{Data with Optimal Density: 3D star coordinates plots for simple clusters.}
 \label{fig:starcoords3d-initial-clusters}
\end{figure*}
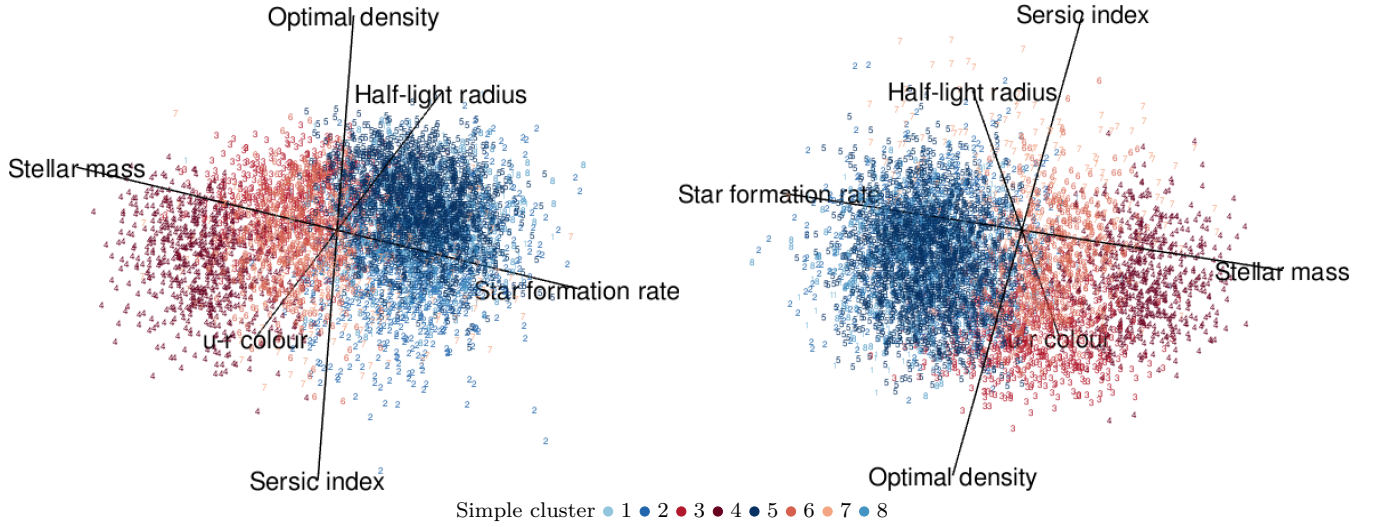

The simple clusters are further illustrated in Fig.~\ref{fig:starcoords3d-initial-clusters} using the 3D visualization framework of \citet{zhuetal21}, where the cluster locations relative to the projected feature directions highlight their main distinguishing characteristics, as summarised in Table~\ref{tab:simple-cluster-interpretation}. The red-sequence galaxies are preferentially located toward higher stellar mass and larger S\'ersic index, while suppressed star formation rates. Conversely, the blue-sequence systems occupy regions characterized by lower stellar mass and S\'ersic index, and exhibit elevated star formation activity. Collectively, the estimated simple clusters demonstrate clear physical and environmental differentiation associated with galaxy formation and evolution, and indicate that massive and red galaxies preferentially inhabit denser environments, while lower-mass, star-forming galaxies are more common in lower-density regions.

\begin{table*}[!ht]
\centering
\caption{Data with Optimal Density: Galaxy counts and astrophysical interpretation of the simple clusters. The descriptions are based on the feature distributions in Fig~\ref{fig:opt-initial-group-feature} and the 3D representations relative to the six feature directions in Fig~\ref{fig:starcoords3d-initial-clusters}.}
\label{tab:simple-cluster-interpretation}
\begin{tabularx}{\textwidth}{c| c| p{0.3\textwidth}| X}
\hline\hline
\bf{Group} & \bf{Galaxy count} & \bf{Approximate population} & \bf{Main distinguishing characteristics} \\
\hline\hline
\bf{\textcolor{gamaopt1}{1}} & $285$  & Low-mass blue/star-forming sequence 
& Low stellar mass, blue colour, relatively high SFR, low S\'ersic index. \\
\hline
\bf{\textcolor{gamaopt2}{2}} & $874$  & Compact low-mass blue/star-forming sequence 
& Low stellar mass, blue colour, high SFR, compact sizes, low environmental density. \\
\hline
\bf{\textcolor{gamaopt3}{3}} & $604$  & Environmentally quenched red sequence 
& Red colour, low SFR, intermediate-to-high stellar mass, highest optimal density. \\
\hline
\bf{\textcolor{gamaopt4}{4}} & $664$  & Mass-quenched red sequence 
& Highest stellar mass, reddest colour, lowest SFR, high S\'ersic index. \\
\hline
\bf{\textcolor{gamaopt5}{5}} & $1,662$ & Extended low- to intermediate-mass blue/star-forming sequence 
& Blue colour, high SFR, larger sizes than other blue groups, low S\'ersic index. \\
\hline
\bf{\textcolor{gamaopt6}{6}} & $496$  & High-mass end of a star-forming sequence 
& High stellar mass, relatively large size, moderate colour, SFR above the quenched groups. \\
\hline
\bf{\textcolor{gamaopt7}{7}} & $209$  & Transition population and/or sources
with large uncertainties 
& Broad feature distributions, largest sizes, high S\'ersic index, intermediate colour and SFR. \\
\hline
\bf{\textcolor{gamaopt8}{8}} & $512$  & Low- to intermediate-mass blue/star-forming systems 
& Blue colour, high SFR, low S\'ersic index, lowest environmental density. \\
\hline
\end{tabularx}
\end{table*}

\begin{table*}[!ht]
\centering
\caption{Data with Optimal Density: Estimated feature means and standard deviations (in parenthesis) for simple clusters.}
\label{tab:gama-feature} 
\addtolength{\tabcolsep}{-0.4em}
\begin{tabular}{c|c|c|c|c|c|c}
\hline\hline
\diagbox[width=7em]{\bf{Cluster}}{\bf{Feature}} 
&\textbf{Stellar mass} & 
\textbf{Star formation rate} &
\bm{$u-r$} \textbf{colour} &
\textbf{Half-light radius} &
\textbf{S\'ersic index} &
\textbf{Optimal density}\\
\hline\hline
\bf{\textcolor{gamaopt1}{1}} 
&$8.77(0.35)$
&$-10.15(1.00)$
&$1.39(0.31)$
&$0.47(0.17)$
&$0.03(0.13)$
&$-0.72(1.90)$
\\
\hline
\bf{\textcolor{gamaopt2}{2}} 
&$8.94(0.44)$
&$-9.59(0.54)$
&$1.32(0.31)$
&$0.21(0.21)$
&$0.26(0.27)$
&$-1.27(1.99)$ 
\\
\hline
\bf{\textcolor{gamaopt3}{3}} 
&$9.67(0.50)$
&$-11.25(0.96)$
&$2.01(0.24)$
&$0.43(0.20)$
&$0.23(0.19)$
&$1.16(1.38)$
\\
\hline
\bf{\textcolor{gamaopt4}{4}} 
&$10.45(0.42)$
&$-12.17(0.72)$
&$2.30(0.16)$
&$0.61(0.29)$
&$0.66(0.14)$
&$0.01(1.92)$ 
\\
\hline
\bf{\textcolor{gamaopt5}{5}} 
&$9.21(0.50)$
&$-9.58(0.35)$
&$1.27(0.20)$
&$0.64(0.16)$
&$0.05(0.15)$
&$-0.97(1.81)$
\\
\hline
\bf{\textcolor{gamaopt6}{6}} 
&$10.39(0.32)$
&$-10.28(0.46)$
&$1.86(0.24)$
&$0.77(0.28)$
&$0.36(0.23)$
&$-0.94(1.83)$
\\ 
\hline
\bf{\textcolor{gamaopt7}{7}} 
&$9.67(0.72)$
&$-10.73(1.17)$
&$1.74(0.43)$
&$1.10(0.61)$
&$0.80(0.26)$
&$-0.39(1.60)$
\\
\hline
\bf{\textcolor{gamaopt8}{8}} 
&$9.07(0.50)$
&$-9.48(0.37)$
&$1.35(0.33)$
&$0.46(0.20)$
&$0.00(0.12)$
&$-1.80(1.92)$
\\ 
\hline
\end{tabular}
\end{table*}

\begin{table*}[!ht]
\centering
\caption{Data with Optimal Density: Estimated factor loadings (in the correlation scale) for simple clusters, along with a heatmap for reference. For clarity
of presentation, values in the interval (-0.1,0.1) are suppressed in the table, but displayed using light colours in the heatmap representation.} 
\label{tab:gama-initial-loadings}
\addtolength{\tabcolsep}{-0.55em}
\setlength\extrarowheight{0.1em}
\begin{tabular}{c|c|c|c|c|c|c|c|c}
\hline
 \hline
\bf{$\bk$}& \bf{$\bq$}& \bf{Stellar mass}& 
\bf{Star formation rate}&
\bf{$\bm{u-r}$ colour}& 
\bf{Half-light radius} & 
\bf{S\'ersic index}& 
\bf{Optimal density}&
\bf{Heatmap}\\
  \hline
 \hline
\multirow{ 1}{*}{\bf{\textcolor{gamaopt1}{1}}} & 1 
&$-0.997$
&$0.368$
&$-0.213$
&$-0.249$
&
&$0.121$
 & \multirow{ 14}{*}
 {
 \includegraphics[height=6.95cm]{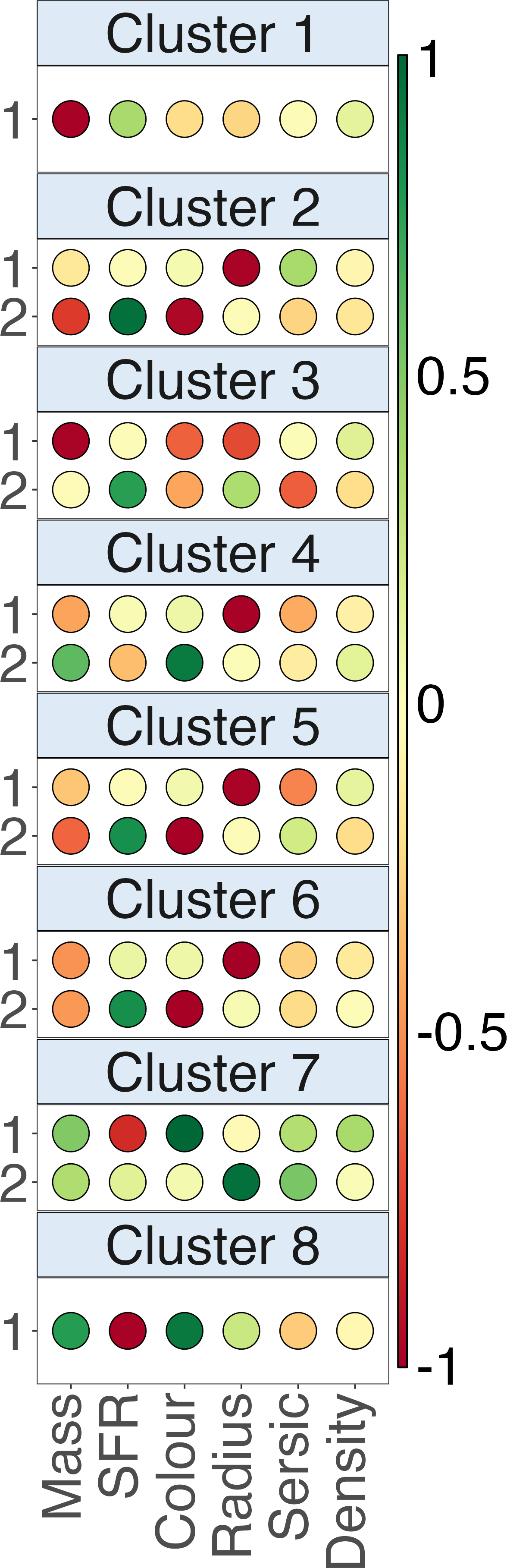}
    }
    \\
 \cline{1-8}
 \multirow{ 2}{*}{\bf{\textcolor{gamaopt2}{2}}} & 1 
 &$-0.158$
&
&
&$-0.995$
&$0.373$
& 
 &\\
 & 2 
 &$-0.776$
&$0.945$
&$-0.974$
&
&$-0.253$
&$-0.159$ 
 &\\
 \cline{1-8}
 \multirow{ 2}{*}{\bf{\textcolor{gamaopt3}{3}}} & 1 
&$-0.993$
&
&$-0.650$
&$-0.724$
&
&$0.149$
 &\\
 & 2 
&
&$0.749$
&$-0.434$
&$0.355$
&$-0.658$
&$-0.210$
 &\\
 \cline{1-8}
 \multirow{ 2}{*}{\bf{\textcolor{gamaopt4}{4}}} & 1 
&$-0.442$
&
&
&$-0.999$
&$-0.418$
&$-0.102$
 &\\
  & 2  
&$0.603$
&$-0.345$
&$0.909$
&
&$-0.128$
&$0.134$
 & \\
 \cline{1-8}
 \multirow{ 2}{*}{\bf{\textcolor{gamaopt5}{5}}} & 1  
&$-0.315$
&
&
&$-0.994$
&$-0.537$
&$0.112$
 & \\
 & 2  
&$-0.639$
&$0.814$
&$-0.995$
&
&$0.217$
&$-0.216$
 & \\
 \cline{1-8}
 \multirow{ 2}{*}{\bf{\textcolor{gamaopt6}{6}}} & 1  
&$-0.492$
&
&
&$-0.996$
&$-0.274$
&$-0.147$
 & \\
 & 2  
&$-0.479$
&$0.817$
&$-0.995$
&
&$-0.219$
&
 & \\
 \cline{1-8}
 \multirow{ 2}{*}{\bf{\textcolor{gamaopt7}{7}}} & 1 
&$0.496$
&$-0.831$
&$0.984$
&
&$0.339$
&$0.370$
 &\\
 & 2 
&$0.346$
&$0.146$
&
&$0.951$
&$0.521$
&
 &\\
 \cline{1-8}
 \multirow{ 1}{*}{\bf{\textcolor{gamaopt8}{8}}} & 1 
&$0.758$
&$-0.997$
&$0.910$
&$0.244$
&$-0.296$
& 
 &\\
 \cline{1-8}
\end{tabular}
\vspace{0.2in}
\end{table*}

\subsubsection{Latent structure analysis}

We further examined the latent structure within each simple cluster using the estimated factor analysers. Table~\ref{tab:gama-initial-loadings} presents both the numerical values and visual representations of the group-wise factor loadings, corresponding to the columns of the loading matrix $\bLambda_k$ described in Section~\ref{sec:fa}. To enhance interpretability, an oblimin rotation \citep{costello2005} was applied to each loading matrix. Only loadings with magnitudes greater than 0.1 are displayed in the table. Each loading value reflects the contribution of an individual feature to a specific latent factor, with the sign ($+$ or $-$) indicating the direction of the relationship between the feature and the factor.

Group~1 is characterised by a single factor that is primarily driven by stellar mass, with additional smaller contributions from half-light radius and $u-r$ colour, which are opposed by smaller to minor components from star formation rate and optimal density. 

In Group~2, the first factor is dominated by half-light radius, with a minor part of stellar mass and small opposing contribution from S\'ersic index. While the second factor is primarily defined by a strong negative loading from $u-r$ colour, accompanied by substantial to minor components from stellar mass, S\'ersic index and optimal density on one side, against a dominant contribution from star formation rate on the other.

For Group~3, the first factor is mainly explained by stellar mass, together with substantial to moderate contributions from half-light radius and $u-r$ colour, opposed by a minor component from optimal density. The second factor contrasts a strong contribution from star formation rate and a moderate part from half-light radius, against opposing loadings from S\'ersic index, $u-r$ colour, and optimal density, whose magnitudes decrease from moderate to smaller to minor.

The first factor in Group 4 is primarily driven by half-light radius, with additional moderate and smaller contributions from stellar mass, S\'ersic index, and optimal density on the same side. The second factor reflects a contrast between strong positive contributions from $u-r$ colour and stellar mass, together with a minor part from optimal density, and opposing smaller to minor components from star formation rate and S\'ersic index.

Group~5 has the first factor dominated by half-light radius, along with moderate to smaller contributions from S\'ersic index and stellar mass, and a minor opposing component from optimal density. The second factor contrasts substantial to small contributions from star formation rate and S\'ersic index, against a dominant part of $u-r$ colour, which is accompanied by moderate to small components of stellar mass and optimal density.

In Group~6, the first factor is mainly explained by half-light radius, together with moderate to minor additional contributions from stellar mass, S\'ersic index and optimal density on the same side. The second factor is dominated by a strong loading from $u-r$ colour, accompanied by moderate to small contributions from the stellar mass and S\'ersic index, which are opposed by a major part of star formation rate.

For Group~7, the first factor contrasts star formation rate against $u-r$ colour, stellar mass, optimal density, and S\'ersic index, whose contributions decrease from dominant to moderate to smaller. The second factor is primarily driven by half-light radius, with an additional moderate contribution from S\'ersic index and smaller to minor components from stellar mass and star formation rate.

Finally, Group~8 has a single factor that is dominated by star formation rate and a smaller part of S\'ersic index, contrasted with strong contributions from $u-r$ colour and stellar mass, together with a smaller component from half-light radius.

We also computed the unbiased estimates of the factor scores $\bF_{i}$ specified in Eq.~\eqref{eq:fascores} using the Bartlett method \citep{bartlett37, hershberger05,distefanoetal09}. The results are given in Table \ref{tab:gama-initial-fascores}, where the mean score values represent the average importance of the latent factors as "rated" by its galaxies members within the group. For Groups~2, 3, 4 and 6, Factor~1 contributes more strongly than Factor~2, while Groups~5 and 7 show the reverse trend. Overall, the distinct factor-loading patterns across the eight simple clusters reveal diverse variability within the galaxy group and reinforce the colour–star-formation dichotomy as a key latent dimension among the clusters.

\begin{table}[H]
\setlength{\tabcolsep}{2pt} 
\setlength\extrarowheight{2.5pt}
\centering
\caption{\label{tab:gama-initial-fascores} Data with Optimal Density: Mean factor scores for simple clusters.}
\begin{tabular}{c|c|c|c|c|c|c|c|c}
\hline\hline
\bf{Cluster} & 
\bf{\textcolor{gamaopt1}{1}} & \bf{\textcolor{gamaopt2}{2}} & \bf{\textcolor{gamaopt3}{3}} & \bf{\textcolor{gamaopt4}{4}} & \bf{\textcolor{gamaopt5}{5}} &
\bf{\textcolor{gamaopt6}{6}} & \bf{\textcolor{gamaopt7}{7}} & \bf{\textcolor{gamaopt8}{8}}\\ 
    \hline\hline
\bf{Factor 1} & $0.07$  &$0.22$ & $0.03$  &$0.01$ &$-0.03$  &$-0.03$ & $0.01$ & $0.01$ \\
\hline
\bf{Factor 2} &--- &$-0.01$  &$0.01$ &$-0.02$  &$0.05$ &$-0.07$  &$0.18$ &--- \\
\hline
\end{tabular}
\end{table}

\subsubsection{Physical picture of the eight simple clusters}

The eight simple clusters are displayed in feature space in Fig.~\ref{fig:opt-initial-group-feature}, and their approximate astrophysical interpretations are summarized in Table~\ref{tab:simple-cluster-interpretation}. Specifically, Groups~3 and 4 stand out as being both red in $u-r$ and with low specific star formation rate (SFR).
These can be associated with quenched populations \citep{peng2010,peng2012}. 
Given the higher values for the environmental measure, Group~3 represents 
environmentally quenched galaxies while Group~4, with log stellar masses $\ga 10$,
represents mass-quenched galaxies \citep{cochrane2018}. 

There are two intermediate clusters. Group~6 has high mass and while it is quite
red, the SFRs are significantly higher than the quenched population. 
This cluster represents the high-mass end of a star-forming sequence \citep{brinchmann04}. 
Group~7 straddles a wide range of physical feature space values. 
This may represent a combination of effects, for example, a transition population and/or 
sources with large uncertainties in feature space. 
The high S\'ersic index and large sizes for some of this space may indicate poor single S\'ersic fits. 
Note this is the smallest cluster in number. 

The four remaining clusters of blue galaxies (Groups 1, 2, 5 and 8) cover low to intermediate
masses of the star-forming sequence. 
The two largest in number of these clusters separate into a compact 
galaxy sample (Group~2) and a more extended galaxy sample (Group~5). 
There is no obvious difference in environmental density between these two groups suggesting this size difference is related to secular evolution. 

The absence of a cluster corresponding to the classical ``green-valley" population is also informative. Galaxies with intermediate colour or star-formation properties are not recovered as a distinct group, but are mainly distributed across Groups~6 and 7. This suggests that, in the present six-dimensional feature space, green-valley galaxies occupy a transitional region between the quenched and star-forming populations.

\subsection{MOBSynC grouping of the M\textit{t}FAD simple clusters}
\label{sec:merge-clusters}
\begin{figure}
    \centering
\includegraphics[width=0.65\linewidth]{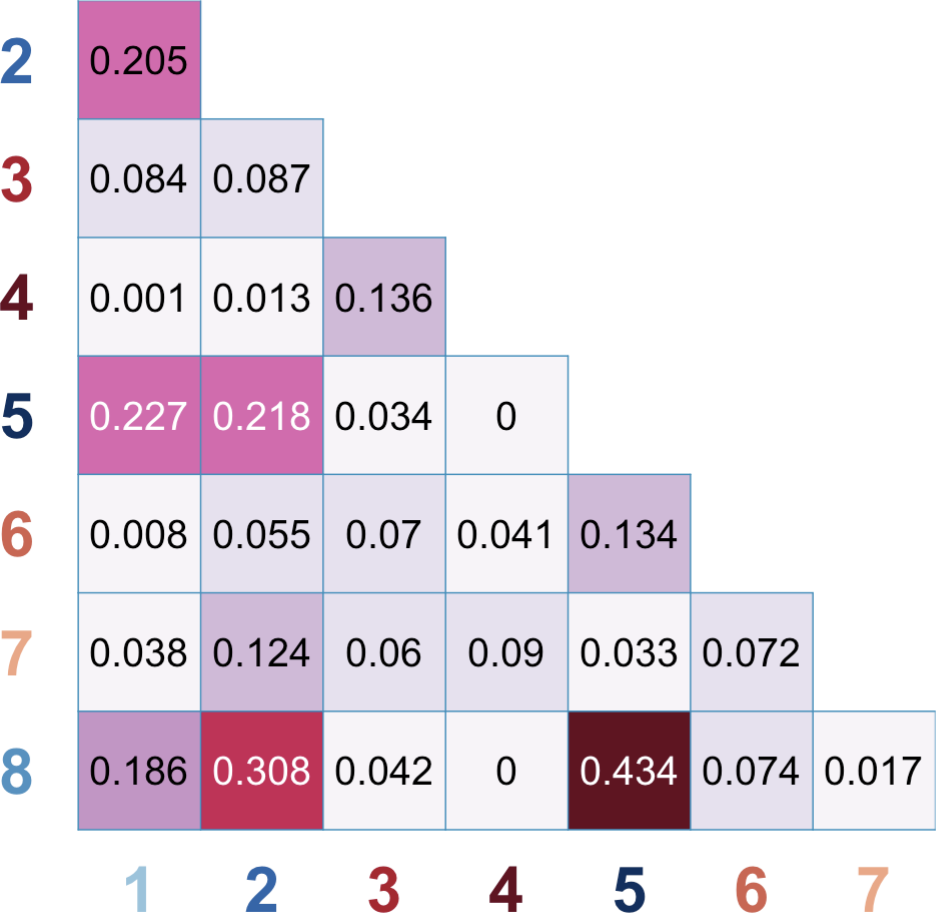}
    \caption{Data with Optimal Density: Pairwise overlap measures between any two of the simple clusters. The generalised overlap is $\ddot{\omega}=0.123$.}
    \label{fig:gama-overlap-initial}
\end{figure}

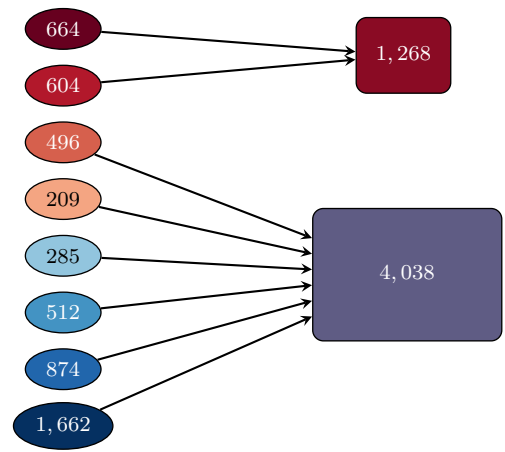
\begin{figure}
\centering
\begin{tikzpicture}
[node distance=2cm]
\node (gopt4) [gopt4,text=white] {$664$};
\node (gopt3) [gopt3, below of=gopt4,yshift=1.25cm,text=white] {$604$};
\node (gopt6) [gopt6, below of=gopt3,yshift=1.25cm,text=white] {$496$};
\node (gopt7) [gopt7, below of=gopt6,yshift=1.25cm,text=black] {$209$};
\node (gopt1) [gopt1, below of=gopt7,yshift=1.25cm,text=black] {$285$};
\node (gopt8) [gopt8, below of=gopt1,yshift=1.25cm,text=white] {$512$};
\node (gopt2) [gopt2, below of=gopt8,yshift=1.25cm,text=white] {$874$};
\node (gopt5) [gopt5, below of=gopt2,yshift=1.25cm,text=white] {$1,662$};
\node[above of=gopt4,font=\bfseries,yshift=-1.25cm] {Initial clustering phase};

\node (gopt34m1) [gopt34m1, right of=gopt4,
yshift=-0.35cm,
xshift=2.5cm,text=white] {$1,268$};

\node (goptno34) [goptno34, below of=gopt34m1,
yshift=-0.9cm,
xshift=0.05cm,text=white] {$4,038$};

\draw [arrow] (gopt1) -- (goptno34);
\draw [arrow] (gopt2) -- (goptno34);
\draw [arrow] (gopt3) -- (gopt34m1);
\draw [arrow] (gopt4) -- (gopt34m1);
\draw [arrow] (gopt5) -- (goptno34);
\draw [arrow] (gopt6) -- (goptno34);
\draw [arrow] (gopt7) -- (goptno34);
\draw [arrow] (gopt8) -- (goptno34);
\node[above of=gopt34m1,font=\bfseries,yshift=-0.92cm] {Final merging phase};
\end{tikzpicture}
\caption{Data with Optimal Density: Flowchart illustrating the application of MOBSynC on simple clusters. Clusters are ordered vertically at each stage according to the average $u-r$ colour of their member galaxies.}
\label{fig:gama-merge}
\end{figure}

Using the simple clusters identified in Section~\ref{sec:simple-clusters}, we further investigated the presence of compound clusters using MOBSynC described in Section~\ref{sec:merge}. Fig~\ref{fig:gama-overlap-initial}
shows the pairwise overlaps among the eight clusters, with a generalised overlap of $\ddot{\omega}=0.123$. Using the selected threshold $\kappa=1$, Groups~3 and 4 merge to form one compound cluster, while the remaining groups combine into another cluster, resulting in two compound clusters at the final stage.
The merging phases and outputs are visualized in Fig~\ref{fig:gama-merge}, where at each phase, the clusters are ordered vertically by average $u-r$ colour so that the reddest to bluest galaxy groups are shown from top to bottom. 

\begin{figure*}
    \centering
   \includegraphics[width = \textwidth]{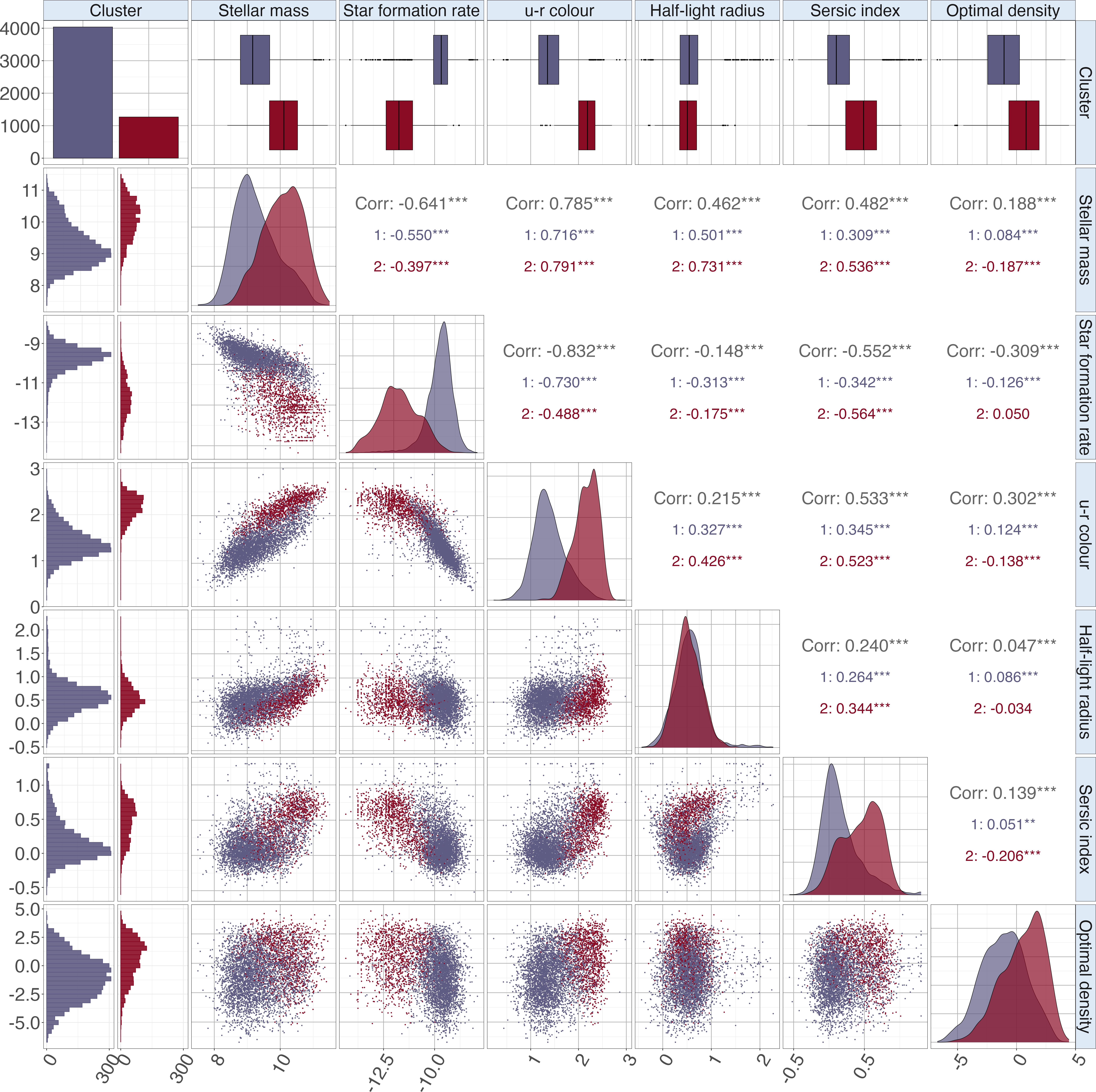}\\
    \centering
    \text{
    Compound clusters
    \tikz\draw[gamaopt.no34,fill=gamaopt.no34] (0,0) circle (.5ex); (1,2,5,6,7,8)
    \tikz\draw[gamaopt.34,fill=gamaopt.34] (0,0) circle (.5ex); (3,4) 
    }
    \caption{Data with Optimal Density: Densities and scatter plots of the six features: Stellar mass, star formation rate, $u-r$ colour, half-light radius, S\'ersic index, and optimal density, for compound clusters (indicated by colours). Correlations between features are shown in the upper panel.}
    \label{fig:final-group-feature}
\end{figure*}

Fig~\ref{fig:final-group-feature}
shows the final feature distributions, demonstrating the two main galaxy populations. Specifically, the compound cluster formed by merging Groups 3 and 4 exhibits substantially higher stellar masses, redder $u-r$ colours, and larger S\'ersic indices and optimal densities, together with a much lower specific star formation rate, compared to the other compound cluster (formed by Groups 1, 2, 5, 6, 7 and 8). Both compound clusters show the largest variation in optimal density, whereas the half-light radius appears to be the least distinctive galaxy property between them.
\begin{figure}
\centering
\includegraphics[width = 0.25\textwidth]{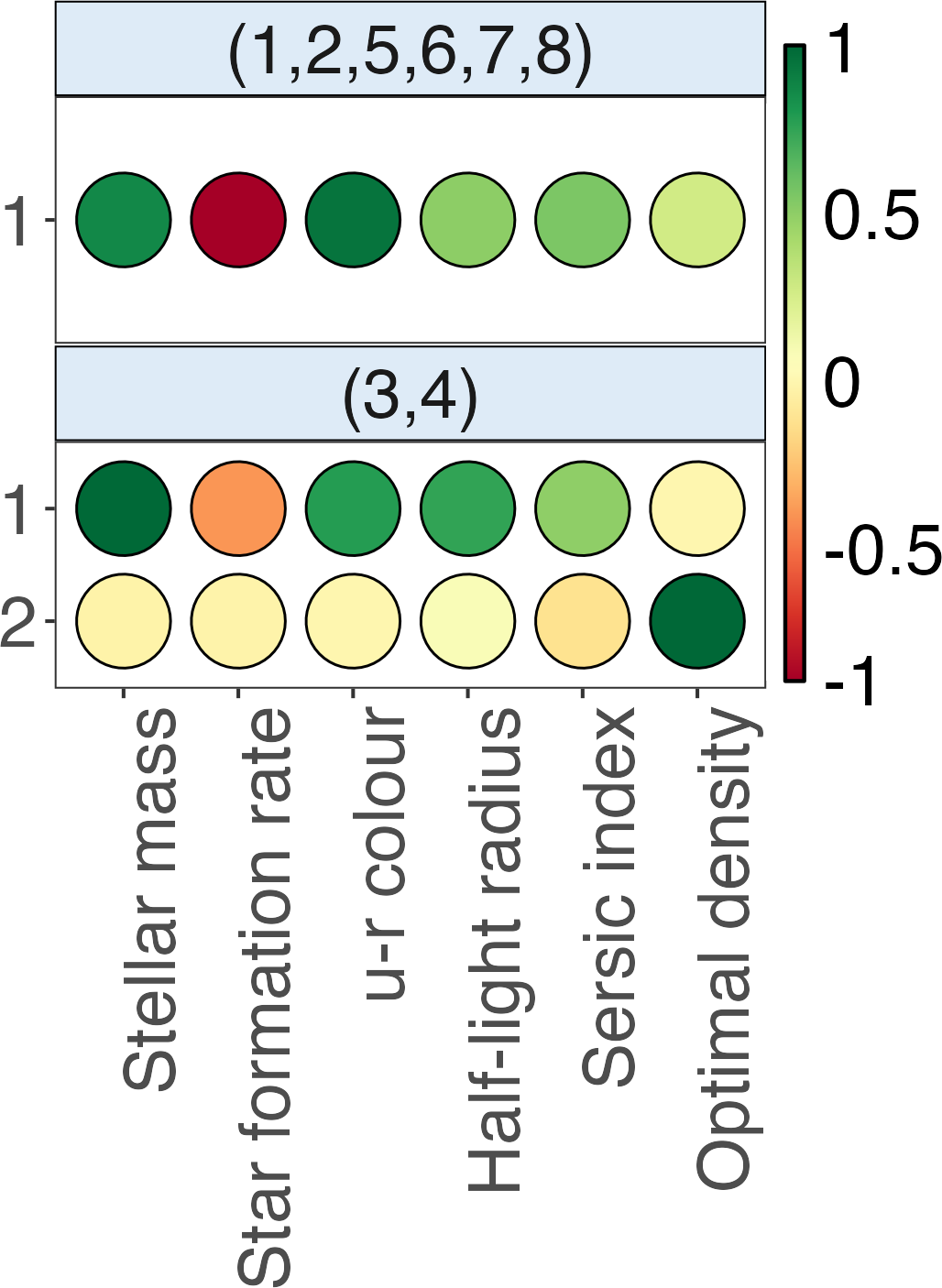}\\
\centering
\caption{Data with Optimal Density: Heatmap of the estimated factor loadings for compound clusters.}
   \label{fig:opt-merge-group-loadings}
\end{figure} 

We further characterised the two compound clusters using a factor model. Because the groups formed by merging simple clusters is no longer normally distributed, we first applied a Gaussian distributional transform (GDT; \citealt{zhuetal21,daiandmaitra24}) to normalize the results, and then performed factor analysis in the Gaussianised space. Fig~\ref{fig:opt-merge-group-loadings}
presents the resulting factor loadings. For the blue sequence (given by the compound cluster formed by merging Groups 1, 2,  5, 6, 7 and 8), the single factor reflects a contrast between star formation rate and strong to moderate to smaller opposing contributions from $u-r$ colour, stellar mass, S\'ersic index, half-light radius, and optimal density in that order. For the red sequence (that is the compound cluster formed by merging Groups 3 and 4), the first factor contrasts star formation rate against stellar mass, $u-r$ colour, half-light radius, and S\'ersic index, whose magnitudes decrease in that order. The second factor is largely dominated by optimal density, opposed by a minor part from S\'ersic index. In sum, MOBSynC clearly distinguishes between the red and blue galaxy systems, consistent with the well-known colour bimodality.

\section{Conclusions}
\label{sec:con}
In this paper, we applied a $t$-mixture of group-specific factor analysers to cluster local-Universe galaxies from the GAMA survey, characterised by five astrophysical features and one combined environmental parameter. We identified eight simple clusters that exhibit distinctivenesses in galaxy properties and in optimal density, revealing valuable insights into the diversity of galaxy populations. These simple clusters provide a finer-level description of the galaxy population, separating, for example, environmentally quenched and mass-quenched red-sequence systems, as well as several star-forming sequence groups that differ in stellar mass, size, morphology, and environment. We further employed MOBSynC to identify larger composite structures by merging less well-separated groups, ultimately yielding two major galaxy classes corresponding to the red and blue sequences. Thus, the familiar red--blue bimodality is recovered as a higher-level structure, while the eight simple clusters reveal additional substructure within this broader bimodality. Each identified cluster was further characterised by latent factors, revealing additional differences in internal galaxy variability and the underlying physical processes.

The two-stage analysis therefore supports a hierarchical interpretation of the local-Universe GAMA galaxy population. At the first level, M$t$FAD identifies simple clusters that capture detailed variation in stellar mass, star formation activity, colour, size, morphology, and environment. At the second level, MOBSynC merges overlapping clusters and recovers the dominant red--blue division. This suggests that the global bimodality remains a major organising feature of local galaxies, but that each side of the bimodality contains physically meaningful substructure associated with quenching pathway, morphology, size, and environment.

From a machine learning perspective, the M$t$FAD clusters are interpreted as model-based probabilistic components in the 6D feature space. This differs from $k$-means clustering, which produces a distance-based partition, and from SOM, which provides a primarily visualization-focused low-dimensional organization of the data. In the present analysis, each M$t$FAD component represents a region of feature space with its own centre, covariance structure, tail behaviour, and latent factor representation. The subsequent MOBSynC analysis then assesses the overlap among these components and merges weakly separated groups to capture the higher-level structure of the feature space.

Although the final galaxy samples excluded incomplete features and outliers, these components can provide useful information on the intrinsic data structure and generative mechanisms. Incorporating measurement errors, missing values, and outliers into statistical modelling therefore remains important. Future work should focus on developing robust, model-based clustering approaches that naturally accommodate these complexities, offering a more complete and realistic characterisation of the data.





\section*{Data Availability Statement}
The galaxy data used in this article comprise local-Universe galaxies from GAMA DR4 \citep{driver22}; available at \url{https://gama-survey.org/dr4/data/cat}. The galaxy features were obtained from the following GAMA DR4 tables: \texttt{MagPhysv06} \citep{driver18}, \texttt{StellarMassesPanChromv24} \citep{taylor11} (for rest-frame $u-r$ colour), and \texttt{BDModelsv05} \citep{casura2022}. The environmental data used in this article consist of three local galaxy environment measures from the GAMA DR4 file server, \texttt{EnvironmentMeasuresv06}, available at \url{https://gama-survey.org/dr4/data/cat/EnvironmentMeasures/v06/}. The processed datasets, together with the \texttt{R} code used to analyse the sample and generate the tables and figures, are publicly available at \url{https://github.com/fanstats/MBC-GAMA}. Software implementing \texttt{M$t$FAD
} will be made publicly available as an \texttt{R} \citep{R} package of the same name, while \texttt{MOBSynC} will be released as part of the publicly available \texttt{SynClustR} package in \texttt{R} \citep{R}.

\bibliographystyle{mnras}
\bibliography{references} 


\appendix

\bsp	
\label{lastpage}
\end{document}